\documentclass{aa}  
\usepackage{graphicx}
\usepackage{txfonts}
\usepackage{amsmath,amssymb}
\usepackage{natbib}
\usepackage[squaren,Gray]{SIunits}
\usepackage{color, colortbl}
\usepackage{array}
\usepackage{longtable,lscape}
\usepackage{multirow}
\usepackage{xcolor}
\usepackage{scrextend}

\usepackage{natbib,twoopt}
\usepackage[breaklinks=true]{hyperref} %% to avoid \citeads line fills

\newcommand {\micron}{\unit{}{\micro\meter}}
\newcommand{\HD}{HD\,181327}

\begin{document} 

   \title{The polarisation properties of the \HD{} debris ring}
   \subtitle{Evidence for sub-micron particles from scattered light observations\thanks{The reduced images (FITS files) presented in Fig \ref{fig_Qphi_Uphi} are only available in electronic form at the CDS via anonymous ftp to \href{http://cdsarc.u-strasbg.fr}{cdsarc.u-strasbg.fr} (130.79.128.5) or via \href{http://cdsweb.u-strasbg.fr/cgi-bin/qcat?J/A+A/}{http://cdsweb.u-strasbg.fr/cgi-bin/qcat?J/A+A/}.}}

   \author{
        J. Milli   \inst{1}
        \and E. Choquet \inst{2}
        \and R. Tazaki \inst{1}
        \and F. M\'enard \inst{1}
        \and J.-C. Augereau \inst{1}
        \and J. Olofsson \inst{4}
        \and P. Th\'ebault \inst{5}
        \and O. Poch \inst{1}
	    \and A.-C. Levasseur-Regourd\inst{6}
	    \and J. Lasue \inst{7}
	    \and J.B. Renard \inst{8}
        \and E. Hadamcik \inst{6}
    	\and C. Baruteau \inst{7}
        \and H.M. Schmid \inst{3}
        \and N. Engler \inst{3}
        \and R. G. van Holstein \inst{9}
        \and E. Zubko \inst{10}
        \and A.M. Lagrange \inst{5,1}
        \and S. Marino \inst{11}
        \and C. Pinte \inst{1}
        \and C. Dominik \inst{12}
        \and A. Boccaletti \inst{5}
        \and M. Langlois \inst{13}
        \and A. Zurlo \inst{14,15,16}
        \and C. Desgrange \inst{1,17}
        \and L. Gluck \inst{1}
        \and D. Mouillet \inst{1}
        \and A. Costille \inst{2}
        \and J.F. Sauvage \inst{2,17}
          }
%        \and G. Chauvin \inst{8,9}
%         \and Q. Kral \inst{7}
%        \and    A.-M. Lagrange \inst{6} 
%         \and C. Perrot \inst{4}
%         \and T. Henning \inst{3}
%         \and M. Min \inst{10,11}
%         \and J.-L. Beuzit \inst{8}
%         \and H. Avenhaus \inst{3}
%        \and A. Bazzon \inst{2}
%        \and T. Moulin \inst{6} 
%        \and M. Llored \inst{12}
%        \and O. Moeller-Nilsson \inst{3}
%        \and R. Roelfsema \inst{13}
%        \and J. Pragt \inst{13}

   \institute{
        Univ. Grenoble Alpes, CNRS, IPAG, F-38000 Grenoble, France\\ %1
         \email{julien.milli@univ-grenoble-alpes.fr}
         \and
          Aix Marseille Universit\'e, CNRS, CNES,  LAM, Marseille, France %2
        \and
        Institute for Particle Physics and Astrophysics, ETH Zurich, Wolfgang-Pauli-Strasse 27, 8093 Zurich, Switzerland %3
%          \and
%        Instituto de F\'isica y Astronom\'ia, Facultad de Ciencias, Universidad de Valpara\'iso, Av. Gran Breta\~na 1111, Playa Ancha, Valpara\'iso, Chile %4
%         \and
%         N\'ucleo Milenio Formaci\'on Planetaria - NPF, Universidad de Valpara\'iso, Av. Gran Breta\~na 1111, Valpara\'iso, Chile %5
          \and 
          Max Planck Institute for Astronomy, K\"onigstuhl 17, D-69117 Heidelberg, Germany  %4
          \and 
          LESIA, Observatoire de Paris, Universit\'e PSL, CNRS, Sorbonne Universit\'e, Univ. Paris Diderot, Sorbonne Paris Cit\'e, 5 place Jules Janssen, 	92195 Meudon, France %5
          \and 
          LATMOS, Sorbonne Universit\'e, CNRS, CNES, Paris, France %6
          \and 
          IRAP, Universit\'e de Toulouse, CNES, CNRS, UPS, Toulouse, France %7
          \and 
          LPC2E, Universit\'e d'Orl\'eans, CNRS, Orl\'eans, France %8
          \and
         European Southern Observatory (ESO), Alonso de C\'ordova 3107, Vitacura, Casilla 19001, Santiago, Chile %9
          \and
          Planetary Atmospheres Group, Institute for Basic Science (IBS), 55 Expo-ro, Yuseong-gu, Daejeon 34126, South Korea %10
          \and 
          University of Exeter, Physics Building, Stocker Road, Exeter EX4 4QL, UK %11
          \and
          Anton Pannekoek Institute for Astronomy, Science Park 904, NL-1098 XH Amsterdam, The Netherlands %12
          \and 
          CRAL, UMR 5574, CNRS, Universit\'e de Lyon, Ecole Normale Sup\'erieure de Lyon, 46 All\'ee d'Italie, F-69364 Lyon Cedex 07, France %13
          \and 
          Millennium Nucleus on Young Exoplanets and their Moons (YEMS), Chile %14
          \and 
          N\'ucleo de Astronom\'ia, Facultad de Ingeniería y Ciencias, Universidad Diego Portales, Av. Ejercito 441, Santiago, Chile %15
          \and 
          Escuela de Ingenier\'ia Industrial, Facultad de Ingeniería y Ciencias, Universidad Diego Portales, Av. Ejercito 441, Santiago, Chile %16
          \and 
           ONERA—Optics Department, 29 Avenue de la Division Leclerc, F-92322 Chatillon Cedex, France %17
             }

   \date{Received 11 September 2023 ; accepted 22 November 2023 }

% \abstract{}{}{}{}{} 
% 5 {} token are mandatory
 
  \abstract
   {
Polarisation is a powerful remote-sensing tool to study the nature of particles scattering the starlight. It is widely used to characterise interplanetary dust particles in the Solar System and increasingly employed to investigate extrasolar dust in debris discs' systems.
   }
   {
   We aim to measure the scattering properties of the dust from the debris ring around \HD{} at near-infrared wavelengths.
   }
   {
   We obtained high-contrast polarimetric images of \HD{} in the H band with the SPHERE / IRDIS instrument on the Very Large Telescope (ESO). We complemented them with archival data from HST / NICMOS in the F110W filter reprocessed in the context of the Archival Legacy Investigations of Circumstellar Environments (ALICE) project. We developed a combined forward-modelling framework to simultaneously retrieve the scattering phase function in polarisation and intensity. 
   }
   {
   We detected the debris disc around \HD{} in polarised light and total intensity. We measured the scattering phase function and the degree of linear polarisation of the dust at 1.6\micron{} in the birth ring. The maximum polarisation is $23.6\% \pm 2.6\%$ and occurs between a scattering angle of $70^\circ$ and $82^\circ$.
   }
   {
   We show that compact spherical particles made of a highly refractive and relatively absorbing material in a differential power-law size distribution of exponent $-3.5$ can simultaneously reproduce the polarimetric and total intensity scattering properties of the dust. This type of material cannot be obtained with a mixture of silicates, amorphous carbon, water ice, and porosity, and requires a more refracting component such as iron-bearing minerals. We reveal a striking analogy between the near-infrared polarisation of comets and that of \HD. The methodology developed here combining VLT/SPHERE and HST/NICMOS may be applicable in the future to combine the polarimetric capabilities of SPHERE with the sensitivity of JWST.
   }

% stress more what makes this new measurement so exciting 


   \maketitle
%
%-------------------------------------------------------------------

    \section{Introduction}

Debris discs are massive analogues of the asteroids and Edgeworth-Kuiper belts in the Solar System \citep[e.g.][for a review]{Hughes2018}. They are made of a population of large kilometre-sized planetesimals, which produce smaller dust particles through mutual collisions. The dust' thermal emission creates an infrared excess in the spectral energy distribution (SED) of the star detectable above the photospheric emission with space-based infrared telescopes. Large infrared surveys revealed that at least 15\% to 30\% of main-sequence stars host a debris disc \citep[e.g. ][]{Matthews2014,Montesinos2016}, and this detection rate can reach 75\% for early-type stars in young moving groups \cite[F stars in the $\beta$ Pic moving group,][]{Pawellek2021}. This means that kilometre-sized planetary embryos are a common outcome of stellar formation. Due to limitations in sensitivity, the detected debris discs have a dust mass several orders of magnitude higher than that in our Solar System at the present time  \citep{Krivov2020}, and they mostly belong to younger systems of a few tens to a few hundreds of millions of years. As collisional activity decays with age \citep{Wyatt2006} and might peak after planet formation, those discs are compatible with our view of the young Solar System after a few hundred million years when it was collisionally very active \citep{Booth2009}. Dust is therefore an important ingredient in these planetary systems and represents a valuable observable that can shed light on the architecture of the underlying planetary system and on the nature of the building blocks of planets to constrain the system formation. Major advances in high-angular resolution and high-contrast imaging techniques now allow one to angularly resolve and isolate the discs from the glare of the central star, opening up new perspectives to characterise the properties of the dust particles through their scattered starlight. About fifty-five debris discs have been detected in scattered light so far, but only a dozen of them have a high enough surface brightness and a favourable geometry or inclination allowing one to extract the scattering properties such as the phase function dependence with the scattering angle. This is the case for Fomalhaut \citep{Min2010}, \HD{} \citep{Stark2014}, HD\,61005 \citep{Olofsson2016}, HR\,4796 \citep{Perrin2015,Milli2017,Milli2019,Olofsson2020,Chen2020,Arriaga2020}, HD\,35841 \citep{Esposito2018}, HD\,191089 \citep{Ren2019} or TWA\, 7 \citep{Ren2021}. They typically exhibit a prominent peak of forward-scattering (HD\,61005, HR\,4796, HD\,35841, or HD\,191089) with a width that varies significantly from one system to the other, and a mild back-scattering, starting either at relatively small scattering angles (from $50^\circ$ for HR\,4796 and maybe even less for Fomalhaut), or at angles larger than $90^\circ$ (HD\,191089 and \HD). 

Even fewer debris discs have been analysed in polarised light, jointly to total intensity, to extract the degree of linear polarisation as a function of the scattering angle (see Table \ref{tab_DoLP_discs}), mostly because obtaining high-fidelity and flux-calibrated images simultaneously in total intensity and linearly polarised light is difficult and possible so far only for the brightest systems. This is the case for the debris disc detected around the star \HD, a young \citep[18.5\,Myrs, ][]{Miret-Roig2020} F5/6V star member of the $\beta$ Pictoris moving group and located at a distance of $48.2 \pm 0.2$ pc \citep{Gaia2018}. Its infrared excess $L_\text{IR,disc}/L_\star$ is estimated at 0.2\% \citep{Lebreton2012}. Hubble Space Telescope (HST) images in the near-infrared from NICMOS \citep[1.1\micron,][]{Schneider2006} and in the optical from ACS (0.6\micron) and later from STIS \citep[0.4-0.8\micron,][]{Stark2014} revealed a wide ring inclined by $28.5^{\circ +2.1^\circ}_{\,-2.0^\circ}$, with an inner edge at $76.6\pm1.0$\,au and a maximum brightness at $84.2\pm1.0$\,au \citep[deprojected semi-major axis values from][taking into account the revised distance to the star]{Stark2014}. The ring is about 25\,au wide, with a cleared interior and  an extended nebulosity detected in the optical up to $\sim400$\,au \citep{Schneider2006}. 
\citet{Stark2014} extracted the scattering phase function (hereafter SPF), showing changes as a function of the distance to the star, compatible with dust segregation in the system. The width of the forward-scattering peak increases with the distance to the star, which suggests that smaller particles dominate the scattered light at larger distances. This is an expected behaviour for a collisionally active ring beyond which should lie a halo of small particles, produced in the main ring and placed on high-eccentricity orbits by radiation pressure. \citep{Lecavelier1996,Strubbe2006,Thebault2008}.
This hypothesis is further supported by millimetre observations with ALMA at 1.3mm \citep{Marino2016}, showing a ring with a semi-major axis at maximum dust density that is $4.2\pm1.1$\,au smaller compared to optical observations (after correcting for the latest distance estimate).

\begin{table}
\caption{Discs with the degree of linear polarisation extracted over a range of scattering angles.}             % title of Table
\label{tab_DoLP_discs}      % is used to refer this table in the text
\centering                          % used for centering table
%\begin{tabular}{p{1cm} p{2cm} p{1cm} p{3cm}}        % centered columns (4 columns)
\begin{tabular}{cccc}        % centered columns (4 columns)
\hline\hline                 % inserts double horizontal lines
Host & Range of & Band & Reference  \\    % table heading 
star & scat. angles &  &   \\    % table heading 
\hline    
HR 4796 & 20$^\circ$-120$^\circ$ & K & \citet{Arriaga2020} \\   
HD 35841 & 22$^\circ$-125$^\circ$ & H & \citet{Esposito2018} \\  
HD 191089 & 30$^\circ$-130$^\circ$ & H & \citet{Ren2019} \\  
HD 114082 & 25$^\circ$-145$^\circ$ & H & \citet{Engler2023} \\  
% HD 117214 & $^\circ$-$^\circ$ &  & \citet{Engler2018} \\  
% HD 61005 & $^\circ$-$^\circ$ &  & \citet{Olofsson2016} \\  
% HD 15115 & $^\circ$-$^\circ$ &  & \citet{Engler2019} \\  
\HD{} & 60$^\circ$-120$^\circ$ & H & this work \\  
\hline                                   %inserts single line
\end{tabular}
\end{table}

A detailed modelling of the SED of the dust constrained by the resolved images of the disc was performed in \citet{Lebreton2012}. It suggests the disc contains micron-sized particles (minimum size 0.9\micron), made of porous amorphous silicates ($\sim12\%$ in volume) and carbonaceous material  ($\sim23\%$) surrounded by an important layer of ice  ($\sim65\%$)	in a  $\sim65\%$ porous structure, possibly fluffy.  

\citet{Schneider2006} and \citet{Stark2014} however revealed some tensions between the minimum grain size suggested by SED modelling of about 1\micron{} and the scattered light behaviour which favours smaller, sub-micron particles.  In an attempt to explore this inconsistency, we present here the first polarimetric observations of this system and extract the degree of linear polarisation. We describe our observations in Sect. \ref{sec_obs}, analyse the morphology in  Sect. \ref{sec_morpho}, extract the phase function and degree of linear polarisation in Sect. \ref{sec_spf} and discuss the scattering properties of the ring in Sect. \ref{sec_modelling} before concluding in Sect. \ref{sec_conclusions}.

%--------------------------------------------------------------------
\section{Observations}
\label{sec_obs}

\subsection{VLT/SPHERE instrumental setup}
\label{sec_ins_setup}

The star \HD{} was observed on the night of 15 May, 2017, with the Spectro-Polarimetric High-contrast Exoplanet REsearch instrument \citep[SPHERE,][]{Beuzit2019}, as part of the Guaranteed Time Observations of the instrument consortium\footnote{ESO programme 099.C-0147(B)}. SPHERE is a high-contrast imager fed by an extreme adaptive optics (AO) system \citep{Sauvage2016_SAXO} to correct for the atmospheric turbulence and static aberrations. We used the near-infrared arm of SPHERE in dual-beam polarimetric imaging \citep[DPI,][]{Langlois2010,vanHolstein2020,deBoer2020}. The IRDIS imager splits the incoming light in two channels, and in DPI, a set of polarisers with orthogonal transmission axes is inserted in the dual-filter wheel in order to simultaneously image the light linearly polarised in two orthogonal directions. IRDIS provides a 11\arcsec$\times$11\arcsec{} field of view with a pixel scale of 12.25 mas \citep{Maire2016}. To benefit from a high-Strehl atmospheric correction in the near infrared, we used the broad-band H filter ($\lambda=1.625$\micron, $\Delta\lambda=0.29$\micron). Observations were carried out in field-stabilised mode, because at the time of observations, the pupil-stabilised mode for DPI was not yet offered \citep{vanHolstein2017}. They used the apodised Lyot coronagraph of radius 92.5\,milliarcsecond (mas) to block the stellar light. 

The details of the coronagraphic data obtained are summarised in Table \ref{tab_log} along with the atmospheric conditions. One polarimetric cycle is made of images recorded at four half-wave plate (HWP) switch angles to measure the Stokes parameters $Q^+$, $Q^-$, $U^+$, and $U^-$. It is referred to as a HWP cycle hereafter and in Table \ref{tab_log}. Unfortunately for the first 15 HWP cycles out of 41 in total, only two positions of the HWP for $Q^+$ and $Q^-$ were recorded.

\begin{table*}
%\begin{minipage}[t]{\columnwidth}
\caption{Log of the SPHERE observations of \HD{} and the reference star HD\, 202917 obtained on the UT date of 16 May, 2017.}
\label{tab_log}
\centering
%\begin{center}
\renewcommand{\footnoterule}{}  % to avoid a line before footnotes
\begin{tabular} {cccccccccc}
%\begin{tabular} {p{2cm}   p{2cm}                       p{1.6cm}                  p{1.6cm}                p{1.4cm}              p{1.2cm}                                   p{1.2cm}    p{1.cm} }
\hline 
UT time &Target & DIT (s) $\times$ & \# HWP  & \# HWP & Seeing\tablefootmark{b} & $\tau_0$\tablefootmark{c} & Wind\tablefootmark{d} & Strehl\tablefootmark{e} & Par. angle\tablefootmark{f} \\
(HH:MM) & &  NDIT\tablefootmark{a}  & positions  & cycles & (") & (ms) & (m/s) & (\%) & start/end ($^\circ$)\\
\hline
\hline
05:59-06:18 & HD\,181327 & $32\times1$    & 2      & 15           &      1.3  &       1.8    &       12.3         & 71  & -58.1/-52.1 \\
06:23-07:34 & HD\,181327  & $32\times2$    & 4    & 15           &      1.2    &       2.4     &       12.5         &  71 & -50.6/-23.7\\
\hline
07:56-09:34 & HD\,202917  & $64\times2$     & 4     & 11           &    1.2     &      1.9           &      11.2       &  58 & -59.5/-23.0 \\
\hline
\end{tabular}
\tablefoot{
\tablefoottext{a}{DIT is the individual Detector Integration Time in seconds and NDIT is the number of DITs.}
\tablefoottext{b}{ from the DIMM.}
\tablefoottext{c}{Coherence time measured by the MASS-DIMM.}
\tablefoottext{d}{Measured 30m above the platform by the Automated Weather Station part of the Astronomical Site Monitoring}
\tablefoottext{e}{Estimate from the Real Time Computer of SPHERE}
\tablefoottext{f}{Parallactic angle}
}
\end{table*}

\begin{figure}
    \centering
   \includegraphics[width=\hsize]{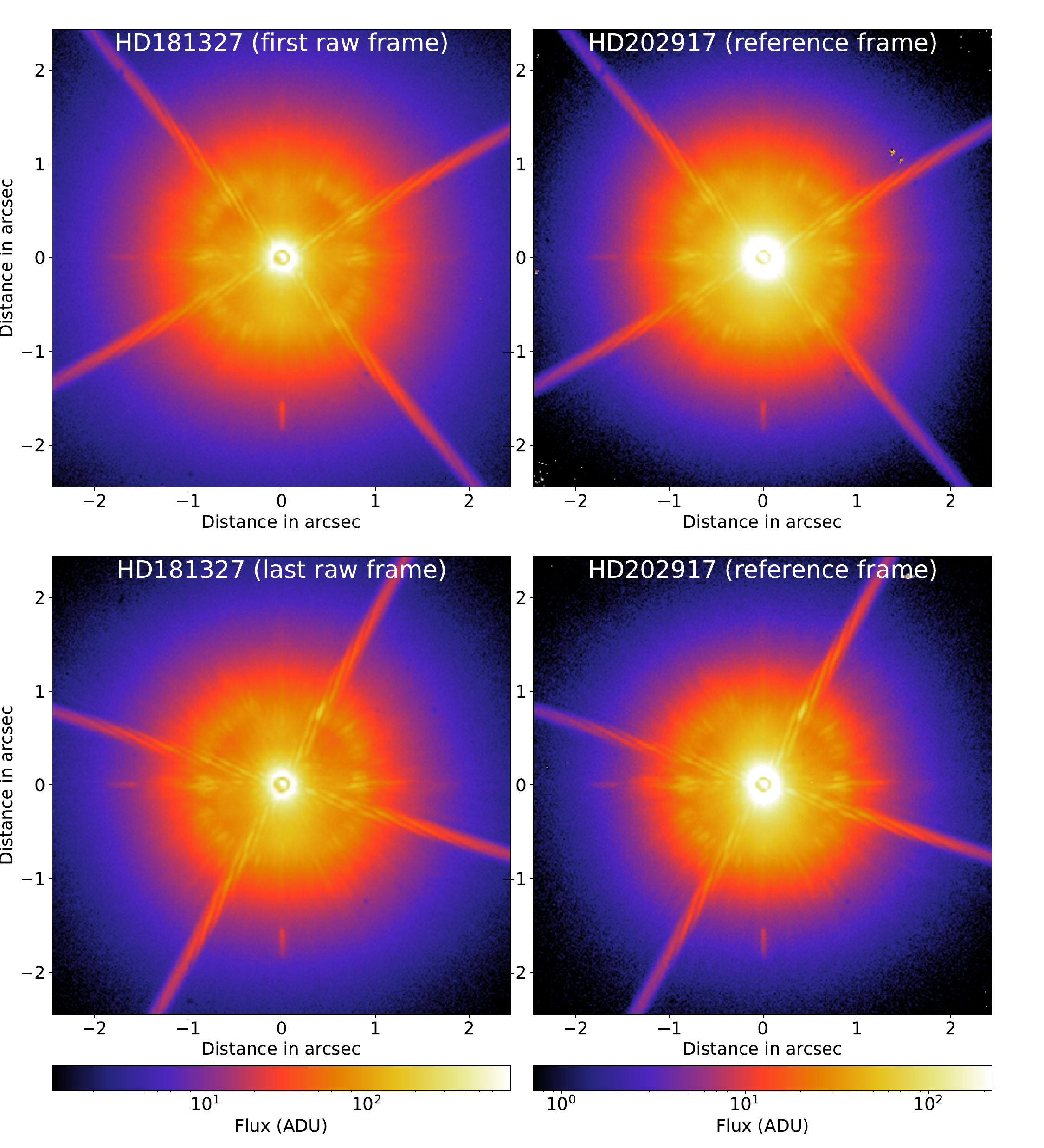}
   \caption{Raw coronagraphic frames of \HD{} and HD\,202917. Left: first (top) and last (bottom) frames of the science target HD\,181327. Right: two examples of frames of the reference star HD\,202917, observed immediately after HD\,181327, and with parallactic angles matching that of the first (top) and last (bottom) raw frame of \HD. North is up and east is to the left. The colour scale is logarithmic and the flux is shown in the raw detector counts called analogue to digital units (ADU).} 
    \label{fig_RDI}
    \end{figure}

Because the observations were field-stabilised, Angular Differential Imaging \citep[ADI,][]{Marois2006} is not applicable to remove the glare of the central star and reveal the disc in total intensity. In addition, ADI is not applicable to objects azimuthally extended such as discs seen under a low inclination, because of severe self-subtraction of the astrophysical signal \citep{Milli2012}. We used an alternative approach by using a reference star observed in the same mode (same filter and coronagraph) and with similar AO correction. This type of data reduction is referred to as Reference Differential Imaging \citep[hereafter RDI, e.g.][]{Ruane2019}. It is commonly applied for space-based imaging \citep[for instance on HST/NICMOS with the ALICE programme,][]{Soummer2014,Choquet2014,Hagan2018}, where the point-spread function (hereafter PSF) is very stable. From the ground, it was also one of the first PSF subtraction technique applied to AO-assisted observations \citep{Mouillet1997} but was later superseded by ADI once pupil-stabilisation was recognised as the most efficient observing strategy to calibrate the halo of speckles of the central star. In field-stabilised observations, the spiders holding the secondary mirror of the telescope are not aligned with the Lyot stop, therefore they diffract the stellar light and relevant reference frames should match the parallactic angles of the telescope in order to have a similar orientation of the spider diffraction pattern. Among all the stars observed in the night of 15 May, 2017, HD\,202917 observed immediately after HD\,181327 provided the best result when used as a reference star to subtract the PSF for the sequence of images of \HD. HD\,202917 was observed along the same range of parallactic angles, under similar atmospheric conditions (see Table \ref{tab_log}), and is only slightly fainter (G magnitude of 8.5 vs 6.9 for \HD). An example showing two raw frames of the science star and two raw frames of the reference star is shown in Fig. \ref{fig_RDI}. This reference star is known to host a faint debris disc, with a low fractional luminosity of $2.5\times 10^{-4}$, only detected in scattered light from space \citep{Schneider2016,Soummer2014} with a very faint peak surface brightness of 0.2mJy/arcsec$^2$ along the semi-minor axis and a semi-major axis of $1.46\arcsec$. It is undetectable in our images neither in the raw frames nor in the reduced polarised or total intensity data. It therefore does not impact our choice of this star as a reference star.

\subsection{Polarimetric data reduction}
\label{sec_pdi_technique}

For each HWP cycle, we obtained the Stokes images $I$, $Q,$ and $U$ and corrected for the instrumental polarization (IP) effects of the complete optical system using the IRDAP pipeline \citep{vanHolstein2020,vanHolstein2020_IRDAP}. The polarisation of the central stellar halo was estimated to $0.08\%\pm 0.07\%$ at the $1\sigma$ level by the pipeline. Therefore the star can be considered to be unpolarised, but we prefer using the image subtracted from this tiny stellar polarisation to have a slightly enhanced data quality.

As we expect the disc polarisation signal to be purely tangential or radial in the case of single-scattering by an optically thin disc illuminated by a single central illumination source, we use the azimuthal Stokes parameter $Q_\phi$ and $U_\phi$ defined as $Q_\phi = -Q\cos(2\phi) - U\sin(2\phi)$ and $U_\phi = +Q\sin(2\phi) - U\cos(2\phi)$  \citep{deBoer2020}, where $\phi$ is the polar angle between the north and the point of interest, measured from the north over east (the position angle). 
$Q_\phi>0$ is equivalent to an azimuthal polarisation component while $Q_\phi<0$ indicates a radial polarisation. The component $U_\phi$ describes the polarisation in the directions $\pm45^\circ$ with respect to the radial direction. 

\begin{figure*}
    \centering
   \includegraphics[width=\hsize]{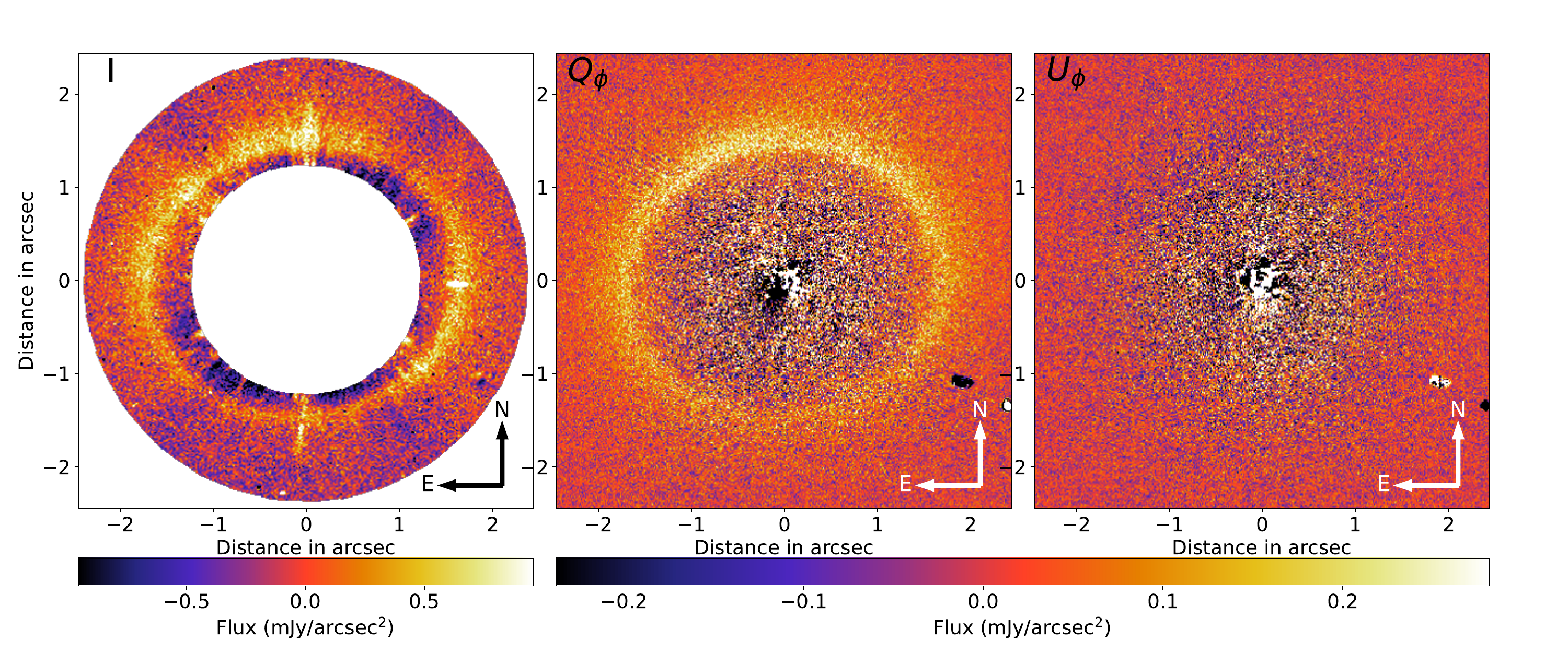}
   \caption{Final images of the intensity (Stokes I, left image), azimuthal Stokes $Q_\phi$ (middle) and $U_\phi$ (right), calibrated in mJy/arcsec$^2$ (linear colour scale). North is up and east to the left. The features in the south-west at $\sim2$\arcsec{} separation are a cluster of bad pixels on the IRDIS detector. A numerical mask of inner radius 1.25\arcsec{} and outer radius and 2.39\arcsec{} was applied on the intensity image.} 
    \label{fig_Qphi_Uphi}
    \end{figure*}

The Stokes $Q_\phi$ and $U_\phi$ are shown in Fig. \ref{fig_Qphi_Uphi} (middle and right image), after conversion to mJy/arcsec$^2$. For the conversion, we estimated the star flux as the total flux of the mean off-axis PSF, encircled in a circular aperture of radius $170$ px (2.1\arcsec), and we took into account the transmission of the neutral density filter and the difference in DIT between the off-axis PSF and the deep coronagraphic images. We assumed a stellar flux density of $4.22\pm0.17 \text{Jy}$ \cite[from 2MASS Johnson H filter,][]{Cutri2003} and a pixel surface area of $12.25\times12.25$ square milliarcseconds.

For the polarimetric data reduction, we did not use the incomplete HWP cycles (first 15 cycles), because the improvement in signal-to-noise (hereafter S/N) was only marginal and we were concerned that combining together $Q$ and $U$ images with different S/N levels might bias the surface brightness distribution of the disc. In total intensity, this issue is irrelevant and all the 150 frames could be combined to form a master cube.  

\subsection{Total intensity data reduction}

As explained in section \ref{sec_ins_setup}, we applied RDI to reveal the total intensity of the disc. The raw frames of the science target \HD{} and its reference star HD\,202917 were flat-fielded, sky-subtracted, bad-pixel corrected and re-centred using the pipeline provided by the SPHERE Data Centre \citep{Delorme2017}. This generated a science data cube consisting of 150 frames for \HD. We further selected the best frames, with no strong speckles and diffraction features at the expected location of the disc, which reduced the data cube to 98 frames. We looked for good references with matching parallactic angles in the data cube of HD\,202917 and selected a reference cube consisting of 88 frames. We divided each of the science and reference cubes in four smaller data cubes sharing a similar range of parallactic angles. We applied for each set of science and reference cubes an RDI algorithm based on Principal Component Analysis \citep[PCA; ][]{Soummer2012,Amara2012}, to remove the quasi-static pattern of the PSF. The residual science images were then stacked together to obtain the image shown in Fig. \ref{fig_Qphi_Uphi} (left). For maximum efficiency, the PCA was optimised in a ring between 1.25\arcsec{} and 2.39\arcsec. The disc is clearly detected at the same location as in the polarised light image (Fig. \ref{fig_Qphi_Uphi} middle), although at a lower S/N because of the difficulty to remove efficiently the stellar halo in RDI compared to PDI. The inner part of the image is not shown because of strong contamination by unsubtracted stellar residuals. The features that extend radially at position angles $0^\circ$, $180^\circ$ and $270^\circ$ are instrumental artefacts not well corrected by the RDI post-processing. They correspond to phase aberrations induced by the pitch of the actuators of the SPHERE deformable mirror at specific spatial frequencies (40 cycles/pupil corresponding to a separation of $1.6\arcsec$ in the H band). On the other hand, the spiders diffraction pattern was well subtracted. 

\subsection{HST/NICMOS archival data}
\label{sec_nicmos}

The HST/NICMOS were originally presented in \citet{Schneider2006}. These observations were originally taken as part of the imaging survey GO-10177 (PI: Schneider), a search for debris discs around 26 targets with strong infrared excess. We reprocessed the deepest coronagraphic sequence obtained with the coronagraphic imaging mode of the NIC2 camera ($0\farcs07565$~pixel$^{-1}$, focal plane mask radius 0\farcs3), in the F110W filter  ($\lambda=1.104$\micron, $\Delta\lambda=0.5915$\micron), obtained at two field orientations in a single spacecraft orbit on 2005 May 2 UT. 
These data were reduced and combined (de-rotated, stacked) using an advanced version of the pipeline developed for the ALICE programme (PI: R. Soummer), a consistent reanalysis of the \textit{HST}/NICMOS coronagraphic archives with advanced starlight subtraction methods \citep{Choquet2014,Hagan2018}, which allowed the discovery of 13 other new debris discs in scattered light \citep{Soummer2014,Choquet2016, Choquet2017,Choquet2018,Marshall2018,Marshall2023}. The final image is shown later in Fig. \ref{fig_NICMOS_image_model_residuals} (left).

\section{Analysis of the morphology and optical depth}
\label{sec_morpho}

The analysis of the morphology and radial profile of the dust density is required to constrain some geometrical properties of the disc and guide us for the extraction of the scattering properties. We used the polarimetric image, free from artefacts inherent to the use of star subtraction techniques in intensity, to investigate those two aspects. 

\subsection{Morphology}
\label{sec_morpho}

\begin{figure}
    \centering
   \includegraphics[width=\hsize]{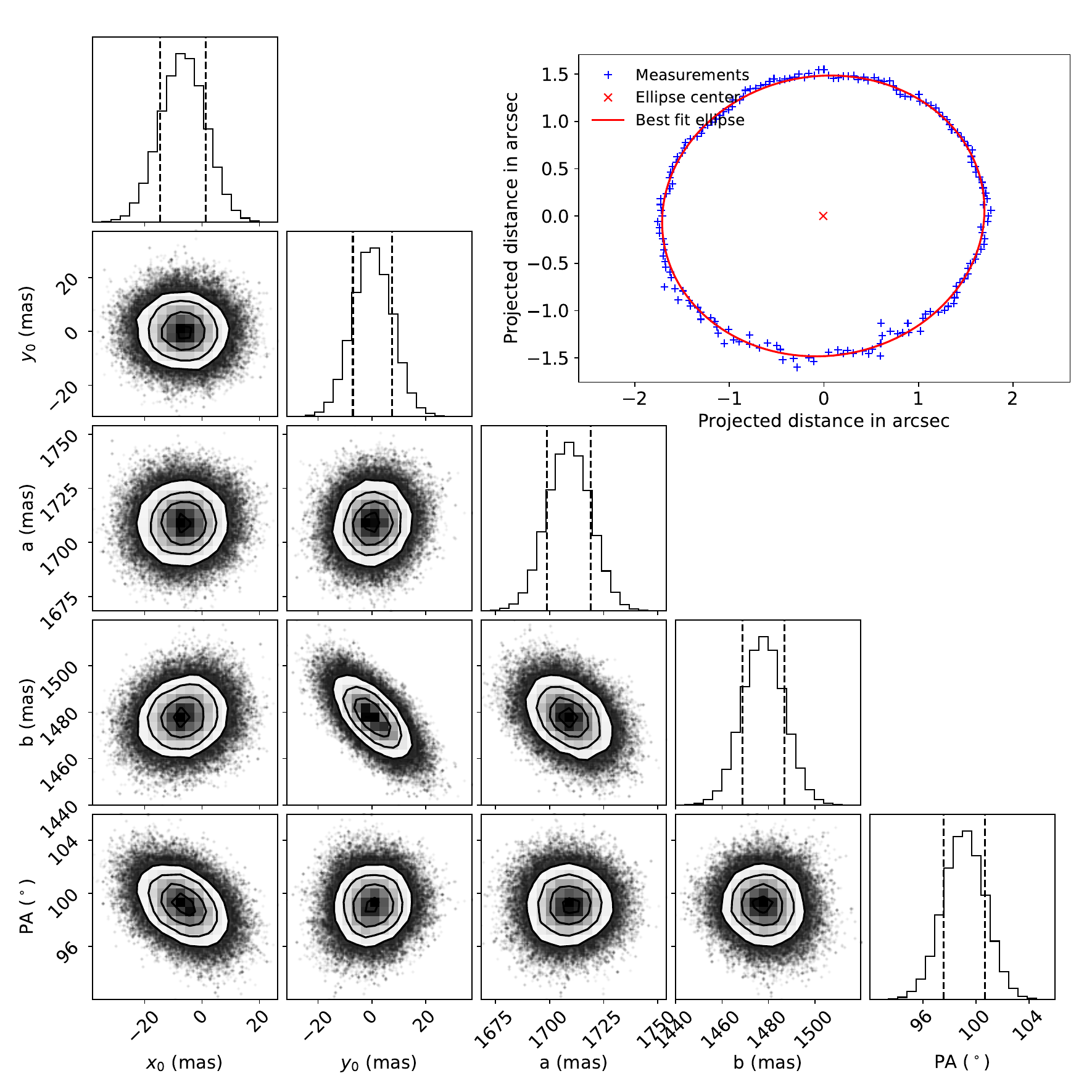}
   \caption{Marginal probability distribution of the parameters of the best ellipse fitting ring: the E-W and N-S offset the ellipse centre $x_0$ and $y_0$, the semi-major and semi-minor axis $a$ and $b$, and the position angle PA. The  graph in the upper right inset shows the data points used as input for the fit in blue, and the best ellipse in red.} 
    \label{fig_MCMC_ellipse}
    \end{figure}

The disc detected around \HD{} appears as an elliptical ring in the sky-projected plane. To derive the morphological parameters of this ring,  we used the polarised intensity image $Q_\phi$ where the ring is detected at a S/N higher than in the total intensity image. We determined the radial location of the disk peak flux and fitted an ellipse to determine elliptical parameters of the dust ring. A similar technique was already used by \citet{Stark2014}. As the disc is found in this study to be non-eccentric, we considered that the location of the belt peak surface brightness reflects the location of the underlying peak dust density  and did not implement the iterative approach to correct for the $1/r^2$ illumination factor.   
 We extracted radial profiles passing through the star and crossing the ring every $2^\circ$ in position angle. For each profile, we determined the radial location of the maximum brightness of the disc by fitting a two-component power law \cite[Eq. 1 in][]{Milli2017}. To find the best ellipse passing through the maximum brightness locations, we implemented the non-linear geometric fitting approach described in \citet{Ray2008} within a Markov Chain Monte Carlo (MCMC) framework \citep{Foreman-Mackey2013} using uniform priors. The result is shown in Fig. \ref{fig_MCMC_ellipse}. The ring has a semi-major axis  $a=1709\pm10$\,mas oriented at a position angle $PA$ of $99.1^\circ\pm1.6^\circ$, a semi-minor axis $b=1478\pm9$\,mas, and its centre is offset by $x_0=-6.8\pm8.2$\,mas and $y_0=0.5\pm7.4$\,mas in right ascension and declination ($x_0<0$ means an east offset, $y_0>0$ means a north offset).      
Using the Kowalsky deprojection technique described in \citet{Smart1930} for binary systems and also applied by \citet{Stark2014,Rodigas2015,Milli2017,Milli2019} on debris discs, we derived the parameters of the true ellipse described by the dust particles in the orbital plane: the true semi-major axis $a$, the eccentricity $e$, the inclination $i$, the argument of pericentre $\omega$ and the longitude of ascending node $\Omega$. The result is given in Table \ref{tab_disc_param}. 
The derived PA and inclination are in agreement with the geometry derived from ALMA \citep{Marino2016,Pawellek2021}. They are also compatible within $1\sigma$ to the parameters derived from the STIS optical images of the ring, except for the semi-major axis, which is found to be $\sim2\%$ larger in the optical.  This small discrepancy could be real: optical and near-infrared images probe different dust populations and \citet{Stark2014} confirmed a spatial segregation of dust particles with smaller particles detected at large separations and larger particles within the birth ring. 
Our analysis also constrains the eccentricity to be below $1.1\%$, hence a poorly constrained argument of pericentre $\omega$. 

\begin{table}
\caption{Deprojected ellipse parameters. The uncertainty is given at $1\sigma$ and contains only the statistical error from the fit and no systematic error from the true north or star registration.}
\label{tab_disc_param}
\renewcommand{\footnoterule}{}  % to avoid a line before footnotes
\centering
\begin{tabular}{c c c}
\hline 
Parameter & IRDIS & STIS\tablefootmark{a} \\
\hline 
$a$ (au) & $82.4 \pm 0.5$ & $84.2  \pm 1.0$\tablefootmark{b} \\
$e$ & $0.004^{+0.007}_{-0.001} $ & $0.02 \pm 0.01$ \\
$i$ ($^\circ$) & $30.2 \pm 1.0$ & $28.5^{+2.1}_{-2.0}$ \\
$\omega$($^\circ$) & $-15^{+75}_{-60}$ & $-70^{+32}_{-33}$ \\
$\Omega$ ($^\circ$) & $99.1 \pm 1.6$ & $11.2 \pm 4.6$\tablefootmark{c}  \\
\hline
\end{tabular}
\tablefoot{
\tablefoottext{a}{\citet{Stark2014}.}
\tablefoottext{b}{This value uses the revised Gaia distance to the star of 48.2\,pc instead of 51.8\,pc from \citet{Stark2014}.}
\tablefoottext{c}{The argument of the ascending node $\Omega$ from \citet{Stark2014} is $90^\circ$ offset from the value derived in this work. This is likely a problem of definition of the origin, taken here as east of north, as the literature mentions various conventions for the deprojection of discs \citep{Chen2020}. }
}
\end{table}

\subsection{Optical depth profile}

\begin{figure}
    \centering
   \includegraphics[width=\hsize]{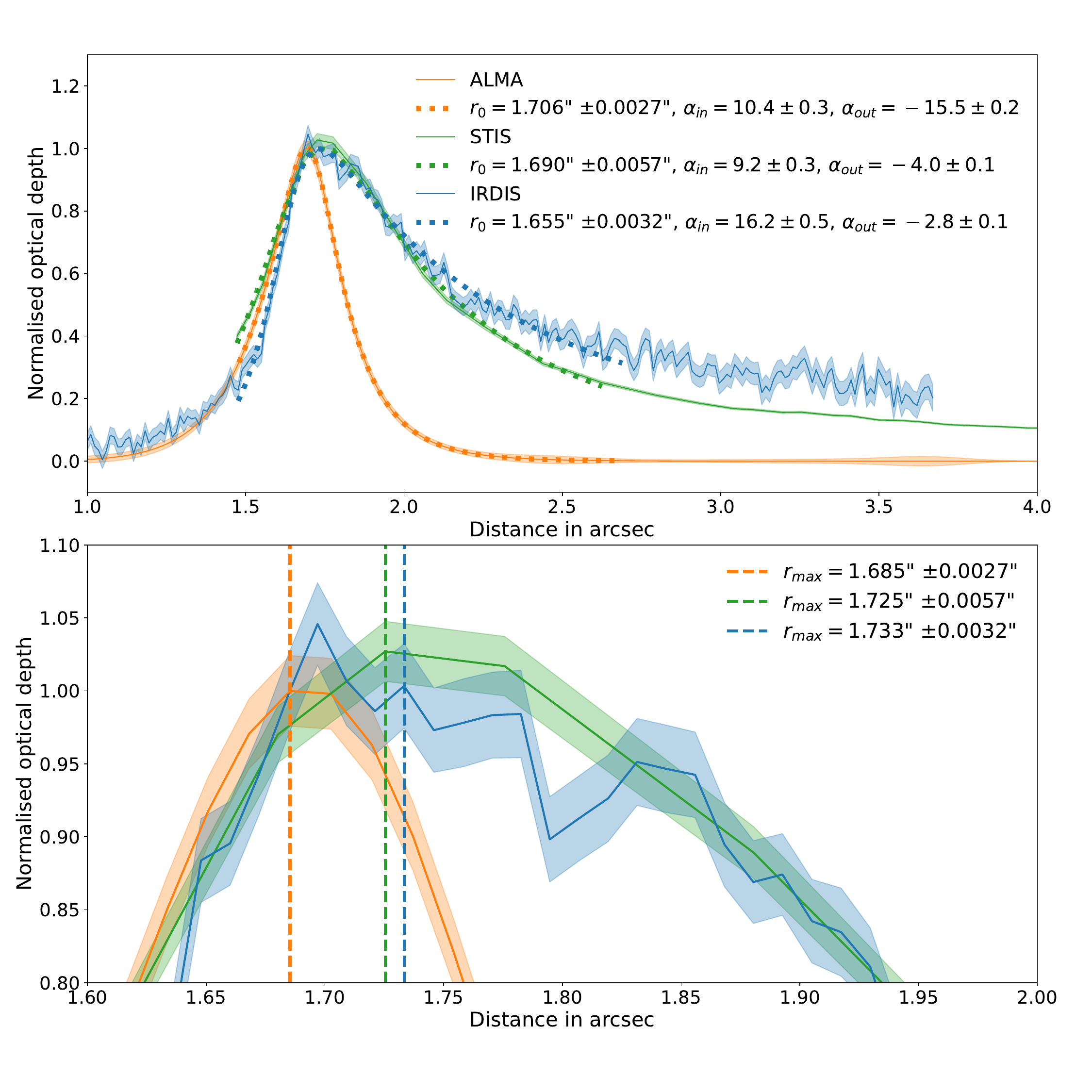}
   \caption{Deprojected normalised optical depth of the disk at mm wavelengths \cite[orange line extracted from data published in][]{Pawellek2021}, in the optical with STIS \cite[green line][assuming a 2\% uncertainty]{Stark2014}, and in the near-infrared with IRDIS (blue line). The top panel shows the data (plain lines) with the $1  \sigma$ uncertainty (shade) and the double-power law fit (dotted line limited to the range of separation where the fit was performed). The bottom panel is a zoom in the region where the optical depth peaks, highlighting with vertical dotted line the location of the maxima.} 
    \label{fig_tau}
    \end{figure}

We followed a methodology similar to  \citet{Stark2014} in order to extract the optical depth $\tau$ of the ring from the polarised intensity IRDIS data. We deprojected the ring and corrected for the $1/r^2$ stellar illumination factor. We measured the median radial profile of the resulting map to estimate the optical depth of the disk. Because the disc brightness varies azimuthally due to the anisotropy of scattering by the dust, we extracted the radial profile in sectors (four wedges centreed respectively on the forward-scattering, backward-scattering and $90^\circ$-scattering sides, with an opening angle of  $90^\circ$ each), normalised to one each sector independently and show a weighted average in Fig. \ref{fig_tau}. 
We note that the optical depth retrieved this way corresponds to the real geometrical optical depth only if all grains have the same scattering properties, which is probably not the case if the system has a large population of very small grains (see Section \ref{sec_discussion_size}), but we use this parameter to allow easy comparison with the STIS and ALMA results. 

To compare with the distribution of mm-sized grains we derive the deprojected intensity profile using the python tool \textsc{frankenstein} \citep{Jennings2020}\footnote{We use an $\alpha$ and $w_{\rm smooth}$ parameters of 1.03 and $10^{-4}$, respectively. The extracted profile and in particular the peak location were not very sensitive to these choices.}, using the ALMA band 7 data ($\lambda=0.88$ mm) and the best-fit disc orientation presented in \cite{Pawellek2021}. The profile is reconstructed from the interferometric visibilities directly, such that the actual resolution is higher than the actual beam size of 0.2\arcsec{} stated in \citet{Pawellek2021}. Subsequently, we multiply the intensity profile by $\sqrt{r}$ to estimate the optical depth at mm wavelengths since the intensity is proportional to the optical depth and $T(r)\propto 1/\sqrt{r}$ at this wavelength. The result is shown in Fig. \ref{fig_tau} (orange line). The radial profile distribution of mm dust peaks at 81~au (1.69\arcsec), consistent with previous estimates for mm-sized grains \citep{Marino2016, Pawellek2021}. We note that \cite{Marino2016} detected extended emission at 1.3\,mm out to 200~au (4.1\arcsec), which is not recovered significantly here due to the lower sensitivity to large-scale structure in these observations compared to those at 1.3mm. 

We then compared the normalised optical depth retrieved from STIS, IRDIS and ALMA by fitting double-power profiles parameterised by the equation 
\begin{equation}
\label{eq_2powerlaw}
\tau(r) = \left( \frac{2}{\left( \frac{r}{r_0} \right)^{-2\alpha_\text{in}} + \left( \frac{r}{r_0} \right)^{-2\alpha_\text{out}} }\right)^{1/2}.
\end{equation}

The best fit parameters along with their $1\sigma$ uncertainty are shown in Fig. \ref{fig_tau} (top panel). The fit was performed between 1.47\arcsec{} and 2.70\arcsec, as shown by the dotted lines. The interest of this global fit, rather than two independent power-laws, is to yield the location of the peak value $r_\text{max}$ (which is usually different from $r_0$ unless $\alpha_\text{out}=\alpha_\text{in}$). A zoom around the maximum value of the optical depth is shown in the bottom panel, with the label giving the location of the peak value $r_\text{max}$.

We note first that the steepness of the inner and outer profiles as well as the location of the maximum $r_\text{max}$ change significantly with the wavelength of observation.
The inner edge profile is steeper in the near-infrared than in the optical, with $\alpha_\text{in,STIS}=9.2 \pm 0.3$ \cite[green dotted line, steeper than the value of 7.0\footnote{\citet{Stark2014} fit a single-power law to the increasing side of the optical depth profile, explaining the difference with the double-power law fit from Eq. \ref{eq_2powerlaw}} reported in][]{Stark2014} and $\alpha_\text{in,IRDIS}=16.2 \pm 0.5$ in the near-infrared with IRDIS (blue dotted line). The STIS and IRDIS instruments do not have the same spatial resolution ($\sim68$ mas sampled by only 1.4 px, vs $\sim45$ mas for IRDIS sampled by 3.7 px) and this difference is likely an effect of the convolution with the instrumental PSF. This could also be physical, as we expect smaller particles to be more sensitive to the Poynting-Robertson drag. 
% ALMA has a $\sim 210$ mas resolution
The comparison between the outer edge IRDIS and STIS profiles show that both profiles are compatible within error bars between 1.72\arcsec{} (maximum optical depth) and 2.2\arcsec. Beyond that, the STIS profile appears steeper than that of IRDIS, but one should take this result with caution because the noise is higher in the near-infrared IRDIS image than in the optical STIS image and the background noise is amplified by the $r^2$ multiplication. This noise already starts to affect the profile beyond 2\arcsec, which may make the outer IRDIS profile less steep, so that the exponent $\alpha_\text{out,IRDIS}=-2.8 \pm 0.1$ is likely too shallow. Indeed the difference with STIS ($\alpha_\text{out,STIS}=-4.0 \pm 0.1$) is surprising because, in the canonical ring+halo scenario, we would expect the outer halo to be mostly populated by grains on high-eccentricity orbits that are bigger than the blow-out size $s_{blow}$. For an F5/6V star, $s_{blow}$ is $\sim 1\micron$ for compact particles and could reach $\sim 5 \micron$ for 75\% porous agglomerates \citep{Arnold2019}, meaning that halo grains should be much larger than $\lambda/2\pi$ for both STIS and IRDIS observations, and we would expect the radial profiles at these two wavelengths to be relatively similar. This similarity might break down if the halo has a significant population of sub-micron grains (see Sect. \ref{sec_discussion_size}), which are poor scatterers in the near IR but not in the optical, but this should lead to a radial profile that is flatter at shorter wavelengths, in contradiction with our results.Another possible reasons for the difference may be the fact that the STIS optical depth profile is estimated from the total intensity image while the IRDIS optical depth profile uses the polarised intensity image. Compact sub-micron particles are expected to scatter more polarised light at near-infrared wavelengths (Rayleigh regime) compared to the optical (Mie regime). In any case, both the STIS and IRDIS profiles are steeper than the canonical $\tau(r) \propto r^{-1.5}$ expected in halos made of $s>s_{blow}$ grains \citep{Strubbe2006}, and a significant population of unbound $s<s_{blow}$ grains would only flatten the profile instead of render it steeper \citep{Thebault2023}. This could be an indication that an additional process is shaping the outer regions of this system.

Interestingly, the mm dust peak density is offset from the micron-size dust peak density. It is located $48 \pm 4.1$ mas ($2.3 \pm 0.2$\,au) closer to the star, suggesting that there is a radial grain size segregation in the parent belt of the star. While not as pronounced as in the halo, such a size segregation is expected within the parent belt, especially for belts having a significant width \citep{Thebault2014}, as in the case for HD181327 \footnote{\cite{Marino2016} found a width of 23\,au in the mm, which corresponds to a relative width of $\sim 25\,\%$}.

In addition, smaller dust particles could be subject to a radial migration because of the presence of gas in the disc, which was indeed detected via $^{{12}}\text{CO}$ (2-1) observations with ALMA \citep{Marino2016}. Gas drag can affect dust grains in different ways. Large particles for which radiation forces are negligible ($s {\gtrsim}100\ \micron$) are typically unaffected by gas drag unless gas densities are large enough in which case they tend to migrate inwards. In contrast, small grains tend to have orbital velocities smaller than the gas and thus gas-drag increases their angular momentum inducing an outward migration \citep{Takeuchi2001}. The end result is that, under the influence of gas, the distribution of small grains can be shifted outwards and concentrated in the outer region of the gas disc, where the gradient in the gas density becomes larger.  Unfortunately, the CO distribution and thus gas radial profile were not well constrained due to insufficient signal-to-noise and thus we cannot quantitatively assess this scenario, which would require significantly more gas from another yet undetected species \citep{Olofsson2022_gas}. 

\section{Extraction and analysis of the scattering properties}
\label{sec_spf}

\subsection{Strategy }
\label{sec_strategy}

Two main families of techniques exist to extract the SPF from discs \citep{Olofsson2020}: direct extraction using aperture photometry \cite[as done for instance in][]{Milli2017,Engler2019,Ren2019}, or forward-modelling with parameterised SPF \cite[as described for instance in][]{Chen2020,Mazoyer2020}. \citet{Olofsson2020} also introduced an non-parametric iterative method in which the phase function is an output of the modelling, but only applicable in polarimetry. In our case, the IRDIS dataset allows us to constrain the polarimetric and total intensity SPF simultaneously from the same observations, reducing considerably the biases and difficulties arising from combining different observations with possibly different instruments, spatial resolutions, observing conditions, etc... However, the type of noise dominating each image is very different, with a polarimetric intensity image dominated by the disc photon noise in the brightest part of the ring and by detector noise in the rest of the image, while the total intensity image is dominated by instrumental diffraction artefacts and other low-spatial frequency noise that the RDI reduction could not eliminate. The SPF direct extraction is therefore possible for the polarimetric image (see section \ref{sec_polar_int}) but too uncertain for the total intensity image. We therefore decided to implement a forward-modelling strategy where we used the NICMOS image of the disc to constrain the SPF in total intensity, the IRDIS polarimetric data set was used to constrain the morphology and polarised SPF, and the IRDIS intensity image was only used to scale the disc model implementing the IRDIS morphology and NICMOS SPF to the correct surface brightness (section \ref{sec_total_frac}). We do not expect significant differences in the SPF between $1.6 \micron$ (IRDIS) and $1.1 \micron$ (NICMOS), hence the choice of this strategy technique.

\subsection{Extraction of the polarised scattering phase function}
\label{sec_polar_int}

To extract the polarised scattering phase function (hereafter pSPF), we started from the polarised intensity image $Q_\phi$  shown in Fig. \ref{fig_Qphi_Uphi} (middle). The $U_\phi$ image shows, as expected, no astrophysical signal, and this image was therefore used to estimate the noise. The noise was assumed to be azimuthally symmetric, and we estimated a 1-$\sigma$ noise map (called $\sigma_{pI}$) and a radial noise profile by computing the root-mean-square (RMS) of the pixel values in 1-px-wide annuli.
We then used two different approaches to extract the polarised SPF from the $Q_\phi$ image: 1) we built a disc model and parametrised the pSPF before iterating to find the best model matching the data, or 2) we extracted the pSPF directly from the data. 

\subsubsection{Forward modelling of the polarised intensity image}
\label{sec_polar_forward_modelling}

To find the best model reproducing the data, we used the python implementation of the GraTeR code \citep{Augereau1999} available as part of the Vortex Imaging Pipeline \cite[VIP\footnote{\label{footnote_vip}A complete description and tutorial of the disc forward-modelling capabilities provided by VIP is available in the online documentation at \href{https://vip.readthedocs.io/en/latest/tutorials/06_fm_disk.html}{https://vip.readthedocs.io/en/latest/tutorials/06\_fm\_disk.html}},][]{Gomez2017}. The complete description of the input parameters is given in Appendix B of \citet{Milli2017} or in the online VIP documentation\footref{footnote_vip}. We set here the inclination $i$ and the longitude of the ascending node $\Omega$ to the values described in Table \ref{tab_disc_param}, and set the eccentricity to zero to avoid degeneracy in $\omega$ and limit the number of free parameters. Regarding the vertical and radial dust density distribution, we set the reference scale height to $\xi_0$=6\,au at the reference radius $r_{0,pI}$ (variable parameter close to the semi-major axis $a$ of $\sim82$ au), we used a Gaussian vertical profile $\gamma=2$, a linear flaring $\beta=1$. A similar parametrisation was used in \citet{Schneider2006}, and these authors constrained the scale height to lie between $4~\text{au} \leq \xi_0 \leq 8~\text{au}$, confirmed later by \citet{Stark2014}, hence our choice to set $\xi_0$ to 6\,au

In a preliminary exploration of the parameter space, we noticed the models favoured a very steep inner edge of the disc, which translates into a large inner slope of the radial dust density distribution $\alpha_{in}$ close to or larger than 15, in agreement with the inner edge slope measured directly in the convolved image and shown in Fig. \ref{fig_tau}. Large values of $\alpha_{in}$ are difficult to constrain due to the limited angular resolution of the image, and discs with $\alpha_{in}$ larger than $15$ would appear the same after convolution with the instrumental PSF. We therefore set this parameter to $15.5$.

We let the outer radial density slopes $a_{out,pI}$ as a variable parameter. We used a custom phase function that we optimised to best reproduce the data. As we want to keep the number of free parameters as low as possible, we observed that the extracted polarised SPF can be well approximated by a cubic interpolation between five points at $0^\circ$, $60^\circ$, $90^\circ$, $120^\circ$  and $180^\circ$ scattering angle (this assumption will be validated later in section \ref{sec_polar_direct_extraction}). Three of those points are kept as free parameters, for scattering angles of $60^\circ$, $90^\circ$  and $120^\circ$ corresponding to the range seen from the Earth. The pSPF at $180^\circ$ and $0^\circ$ are set to zero and twice the value at $60^\circ$ respectively, to provide a smooth SPF between  $60^\circ$ and $120^\circ$ but the value has no physical meaning because it is not probed from the Earth, and it has no impact on the result of the fit.

In total there are five free parameters for the fit:  the outer slope $a_{out,pI}$, the reference radius $r_{0,pI}$, and the 3 interpolation points of the polarised SPF called here $s_{60,pI}$, $s_{90,pI}$, $s_{120,pI}$ that combine the polarised SPF interpolated at $60^\circ$, $90^\circ$  and $120^\circ$ respectively and the disc total scattering cross-section to avoid introducing an additional scaling factor to match the brightness of the disc. The subscript '$_{pI}$' refers to polarised intensity, and is used to distinguish with the total intensity parameters using the subscript '$_I$'.

Rigorously, one should construct synthetic Stokes $Q$ and $U$ images from the ray-traced image of a disc model, assuming purely tangential linear polarisation, then convolve those $Q$ and $U$ images with the measured IRDIS PSF, and reconstruct a synthetic $Q_\phi$ image to be compared to the data. This guarantees the effect of the convolution is correctly accounted for \cite[see Appendix A of][for a detailed description of that procedure]{Engler2018}. In our case, the disc is extended and shows a cavity much larger than the PSF size, therefore one can directly compare the data with the convolved ray-traced image from the disc model. We therefore implemented this strategy, which is faster, after checking that the difference is negligible with the slower but more rigorous approach\footnote{The relative change between the $Q_\phi$ image computed with both methods is less than 0.6\% in the area of the image where there is detectable disc signal. This is negligible with respect to uncertainty associated with other sources of noise, such as the variability of the PSF during the observations.}. We iterated on the five free parameters to derive the best model explaining the data, by minimising the chi squared $\chi_{pI}^2$. It is defined as
\begin{equation}
\chi_{pI}^2 = \displaystyle\sum_{i}    \left[ \frac{Q_\phi-M_{pI}(s_{60,pI},s_{90,pI},s_{120,pI},a_{out,pI},r_{0,pI})}{\sigma_{pI}} \right]^2,
\end{equation}
where $M_{pI}$ is the disc model in polarised intensity, $\sigma_{pI}$ is the noise map, and $i$ is the index of the pixels used to compute the sum. A mask was created to encompass only the pixels between 1.3\arcsec and 2.2\arcsec where disc signal is detected.

\begin{figure*}
    \centering
   \includegraphics[width=\hsize]{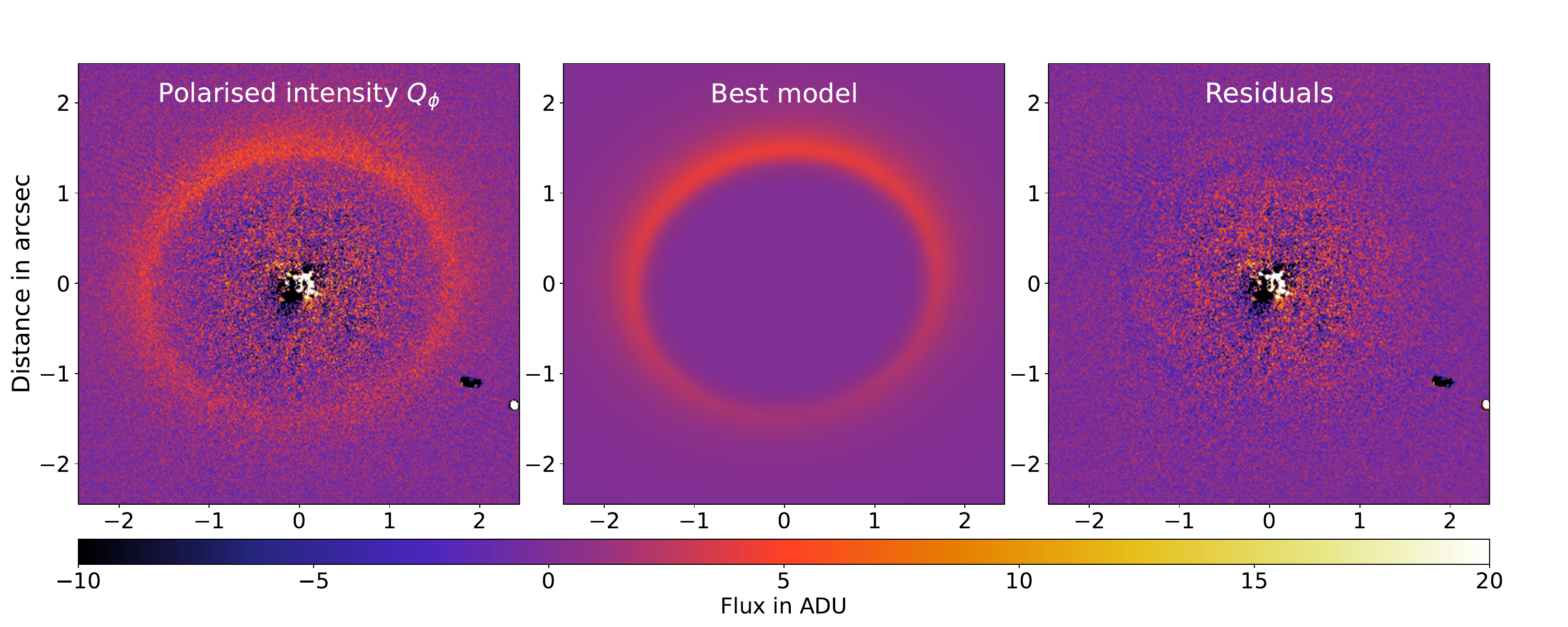}
   \caption{Interpretation of the IRDIS polarised intensity. Left: polarised intensity $Q_\phi$. Middle: best scattered light model explaining the data (unconvolved). Right: residuals after subtraction of the best model.} 
    \label{fig_polar_disc_model}
    \end{figure*}

We used an MCMC with 100 walkers and 500 steps. The chains converged, we considered the first 20\%  of the steps as burn-in phase. 
 The results from the optimisation are summarised in Table \ref{tab_results_mcmc} (column Best fit $\chi_{pI}^2$). The full posterior probability distribution of the free parameters is available in Appendix \ref{fig_cornerplot_PDI}, Fig. \ref{fig_cornerplot_PDI}. The best disc polarised intensity model is shown in Fig. \ref{fig_polar_disc_model} (middle), the reduced $\chi^2$ is 0.76 indicating a good match to the data, as visible in the residual image after subtraction of the best model Fig. \ref{fig_polar_disc_model} right). We reproduced in Fig. \ref{fig_polar_best_SPF} the best scattering phase function, as well as 100 models randomly drawn from the posterior distribution of the parameters $s_{60,pI}$, $s_{90,pI}$ and $s_{120,pI}$.

\begin{table}
\caption{Parametrisation of the disc in polarisation. For each parameter, a uniform prior in the range stated in the second column was used in the MCMC. The error is stated at $5 \sigma$.}
\label{tab_results_mcmc}
\centering
\begin{tabular}{c c c}
\hline 
Parameter & Range & Best fit value \\
\hline 
$s_{60,pI}$ & $[0,7]$  & $3.42 \pm 0.48$ \\
$s_{90,pI}$  & $[0,5]$ & $2.30 \pm 0.32$ \\
$s_{120,pI}$ & $[0,3]$ & $1.33 \pm 0.34$ \\
$a_{out,pI}$ & $[-9,-1]$ & $ -4.32 \pm 0.95$ \\
$r_{0,pI}$ (au) & $[65,100]$ & $79.9 \pm 1.4$ \\
\hline
\end{tabular}
\end{table}

\begin{table*}
\caption{ Summary of the free parameters (first column) and the interval over which a uniform prior was assumed (second column) and their most likely value for each of the three fits performed, to minimise the residuals in polarised intensity with IRDIS (third column), in total intensity with NICMOS (fourth column) or the combination of polarised intensity with IRDIS, and total intensity with NICMOS and IRDIS (fifth column). The uncertainty is stated at $5 \sigma$.}
\label{tab_results_mcmc}
\centering
\begin{tabular}{c c c c c}
\hline 
Parameter & Range & Best fit $\chi_{pI}^2$ & Best fit $\chi_{I,NIC}^2$ & Best fit $\chi^2_{sum}$   \\
\hline 
$s_{60,pI}$ & $[0,7]$  & $3.42 \pm 0.48$ & & $3.52 \pm 0.50$\\
$s_{90,pI}$  & $[0,5]$ & $2.30 \pm 0.32$  & & $2.34 \pm 0.30$ \\
$s_{120,pI}$ & $[0,3]$ & $1.33 \pm 0.34$  & & $1.34 \pm 0.29$ \\
$a_{out,pI}$ & $[-9,-1]$ & $ -4.32 \pm 0.95$  & & $-4.59 \pm 1.07$ \\
$r_{0,pI}$ (au) & $[65,100]$ & $79.9 \pm 1.4$  & & $80.02 \pm 1.08$\\

$s_{60,I}$ & $[0,7]$  & & $4.66 \pm 0.37$  & $4.47 \pm 0.33$ \\
$s_{90,I}$  & $[0,5]$ & & $2.86 \pm 0.22$  & $3.00 \pm 0.21$ \\
$s_{120,I}$ & $[0,3]$ &  &$2.40 \pm 0.30$  & $2.24 \pm 0.25$ \\
$a_{out,I}$ & $[-9,-1]$ &  &$  -7.37 \pm 0.65$  & $-7.39 \pm 0.58$ \\
$r_{0,I}$ (au) & $[65,100]$ & & $81.81 \pm 0.83$ & $81.80 \pm 0.85$ \\

$\Lambda$ & [0,1.8] & & & $0.84 \pm 0.09$ \\
\hline
\end{tabular}
\end{table*}
%
%s60I = $4.47 \pm 0.33$
%s90I = $3.00 \pm 0.21$
%s120I = $2.24 \pm 0.25$
%r0I = $81.80 \pm 0.85$
%aoutI = $-7.39 \pm 0.58$
%s60pI = $3.52 \pm 0.50$
%s90pI = $2.34 \pm 0.30$
%s120pI = $1.34 \pm 0.29$
%r0pI = $80.02 \pm 1.08$
%aoutpI = $-4.59 \pm 1.07$
%scaling = $0.84 \pm 0.09$

\subsubsection{Direct extraction}
\label{sec_polar_direct_extraction}

\begin{figure}
    \centering
   \includegraphics[width=\hsize]{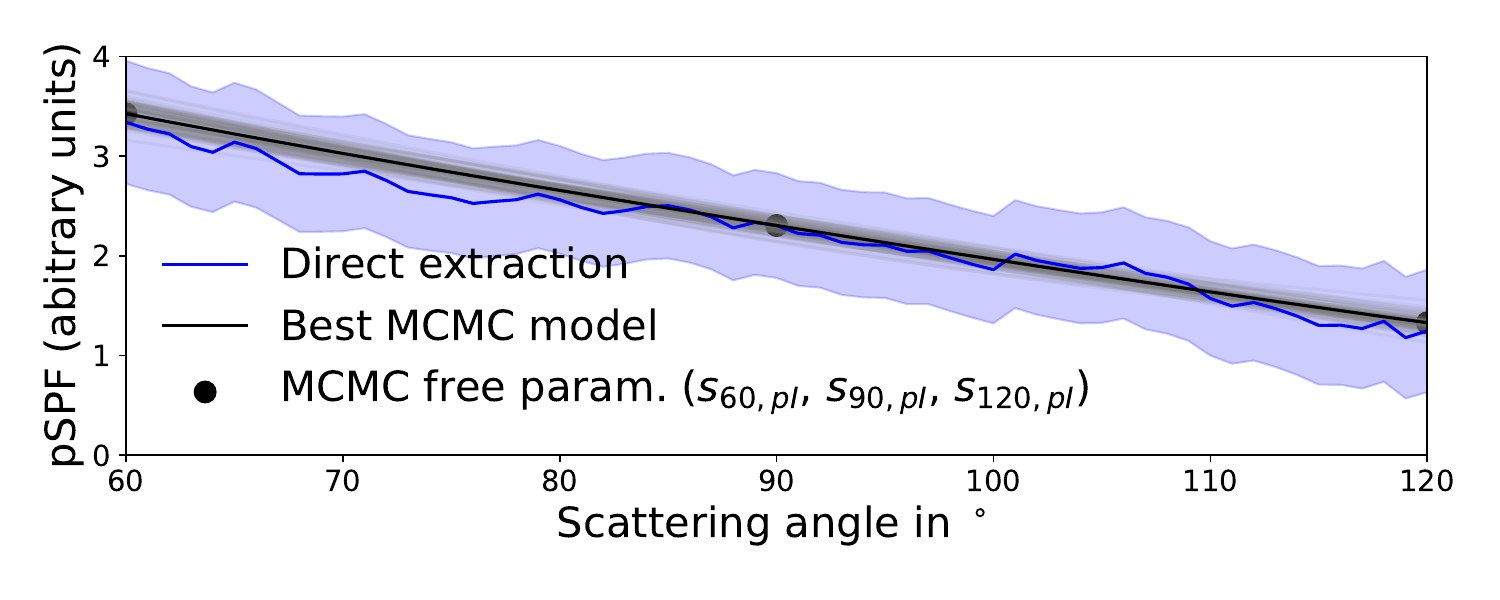}
   \caption{Polarised SPF estimated using the direct extraction method (blue curve) and the parametric model (black curve). The 1-$\sigma$ uncertainty is shown in blue and black shade respectively. Both curves have been normalised to the same value at a scattering angle of $90^\circ$ to allow a meaningful comparison} 
    \label{fig_polar_best_SPF}
    \end{figure}

To avoid depending on the parametrisation of a disc model, the SPF can also be estimated directly from aperture photometry after accounting for various effects. We implemented this alternative approach as a sanitary check and followed the methodology detailed in \citet{Milli2017}.
First, we regularly sampled the best ellipse (as defined in Table \ref{tab_disc_param}). The spacing between each point was set to one resolution element. We associated to each point in the plane of the sky a unique scattering phase angle assuming a thin disk with zero scale height. We performed aperture photometry in elliptical apertures to account for the inclination of the system. The relative effect of the convolution and the offset from the star has no impact within our error bars. The limb brightening effect described in \citet{Olofsson2020}, that can create brightness enhancements in the ansae of a disc has a negligible impact here because of the low inclination of HD\,181327, we therefore did not implement any correction for that. The flux encircled in the aperture is therefore directly proportional to the polarised SPF of the disc. It is overplot in Fig. \ref{fig_polar_best_SPF} (blue line) after normalising it to the same value as the pSPF extracted from the model-based approach (black line). Both approaches agree within error bars, which validates our extraction procedures. The direct extraction yields higher uncertainty because it is a non-parametric approach. 

The pSPF varies linearly with the scattering angle between $60^\circ$ and $120^\circ$. This linear decrease is also visible in the pSPF of HR\,4796 \citep{Arriaga2020} in the H and K band with a similar slope.

\subsection{Total intensity}
\label{sec_total_frac}

A direct extraction of the SPF in total intensity is not possible from the IRDIS total intensity image alone. This image suffers from various artefacts (AO features and low-frequency noise left behind by the RDI reduction) varying azimuthally along the ring that bias any measurement. We illustrate that problem in Appendix \ref{App_SPF_IRDIS}. The model-based SPF retrieval technique is also not applicable here, because the retrieved error bars are too large to allow a meaningful interpretation. This image however still contain some important information on the overall surface brightness of the disc in total intensity, that is key in order to calibrate the polarised intensity and extract the degree of linear polarisation. 

We therefore used the NICMOS image at a slightly different wavelength ($1.104\micron$ vs $1.625\micron$ for IRDIS) to model the SPF. The model used is the same as that used in polarised intensity, with five free parameters: the outer slope $a_{out,I}$, the reference radius $r_{0,I}$, and the 3 interpolation points of the SPF called $s_{60,I}$, $s_{90,I}$, $s_{120,I}$. We decided to allow the reference radius and the outer slope to be different from the best fit values found in polarised light $a_{out,pI}$ and $r_{0,pI}$, because we found that NICMOS favoured a steeper outer edge than IRDIS implying also a slightly larger reference radius ($a_{out}$ and $r_{0}$ are correlated), and our focus is here on the SPF that can be best constrained if the morphology is also well described. 

The effect of the PCA data reduction was taken into account through the following steps: for each disc on-sky orientation in the raw images, the disc model is convolved with the instrumental PSF, projected onto the Karhunen-Lo\`eve modes, and the resulting images are de-rotated and combined to create a forward model called $\text{FM}_\text{I,{NIC}}$ that can be directly subtracted from the NICMOS image $I_\text{NIC}$ to test the quality of the model. We used a similar metric as for IRDIS to estimate the quality of a model: the chi squared $\chi_\text{I,NIC}^2$ computed in a circular annulus where the disc signal is detected:
\begin{equation}
\chi_\text{I,NIC}^2 = \displaystyle\sum_{i}    \left[ \frac{I_\text{NIC}-\text{FM}_\text{I,NIC}(s_{60,I},s_{90,I},s_{120,I},a_{out,I},r_{0,I})}{\sigma_\text{I,NIC}} \right]^2,
\end{equation}
where $\sigma_\text{I,NIC}$ is the NICMOS noise map. This noise map was estimated by processing the reference star images from the NICMOS PSF libraries with the same method and reduction parameters as the NICMOS science images of \HD, as explained in \citet{Choquet2018}. The PSF-subtracted libraries were then partitioned into sets with the same number of frames as \HD, rotated with the target image orientations, and combined. The noise maps were computed from the pixel-wise standard deviation across these sets of processed reference star images.

We implemented again a maximum likelihood approach, using an MCMC framework. The results from the optimisation are summarised in Table \ref{tab_results_mcmc} (column Best fit $\chi_{I,NIC}^2$). The full posterior probabilities are shown in Appendix \ref{App_SPF_NICMOS} and we show in Fig. \ref{fig_NICMOS_SPF} the NICMOS image, the best model and the residuals after subtraction of the forward model. The SPF parametrising the best model is reproduced in Fig. \ref{fig_NICMOS_SPF}, as well as 100 models randomly drawn from the posterior distribution of the parameters $s_{60,I}$, $s_{90,I}$ and $s_{120,I}$. The SPF is monotonously decreasing with scattering angle and seems to flatten to large angles. No backward scattering is detected beyond $110^\circ$, as visible in the optical \citep{Stark2014} for some orbital distances. However we note that the SPF is similar to the SPF measured in the optical  at a separation of 105\,au \cite[Fig. 7 top panel in ][corresponding to a radius of 97.7\,au with the revised Gaia distance]{Stark2014}. Inspecting our residual image (Fig. \ref{fig_NICMOS_image_model_residuals} right) and given the SPF uncertainty (Fig. \ref{fig_NICMOS_SPF}), our NIR observations support the absence of backward-scattering at scattering angles less than $120^\circ$.

\begin{figure}
    \centering
   \includegraphics[width=\hsize]{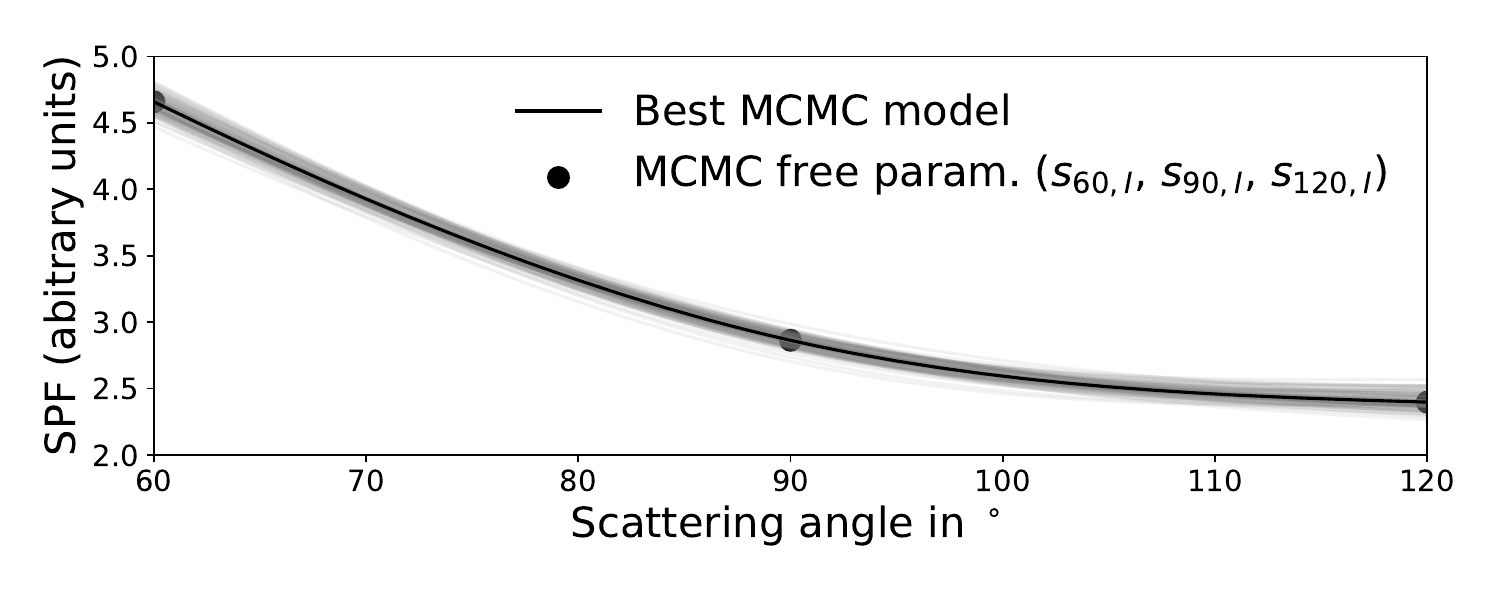}
   \caption{SPF in intensity parametrising the dust scattered light measured by NICMOS the best. The 3 degrees of freedom of the SPF are shown as black dots, and the model performs a cubic interpolation (black curve). } 
    \label{fig_NICMOS_SPF}
    \end{figure}

%\begin{table}
%\caption{Parametrisation of the disc in total intensity. For each parameter, a uniform prior in the range stated in the second column was used in the MCMC. The error is stated at $5 \sigma$.}
%\label{tab_nicmos_mcmc}
%\centering
%\begin{tabular}{c c c}
%\hline 
%Parameter & Range & Best fit value \\
%\hline 
%$s_{60,I}$ & $[0,7]$  & $4.66 \pm 0.37$ \\
%$s_{90,I}$  & $[0,5]$ & $2.86 \pm 0.22$ \\
%$s_{120,I}$ & $[0,3]$ & $2.40 \pm 0.30$ \\
%$a_{out,I}$ & $[-9,-1]$ & $  -7.37 \pm 0.65$ \\
%$r_{0,I}$ (au) & $[65,100]$ & $81.81 \pm 0.83$ \\
%\hline
%\end{tabular}
%\end{table}

\begin{figure*}
    \centering
   \includegraphics[width=\hsize]{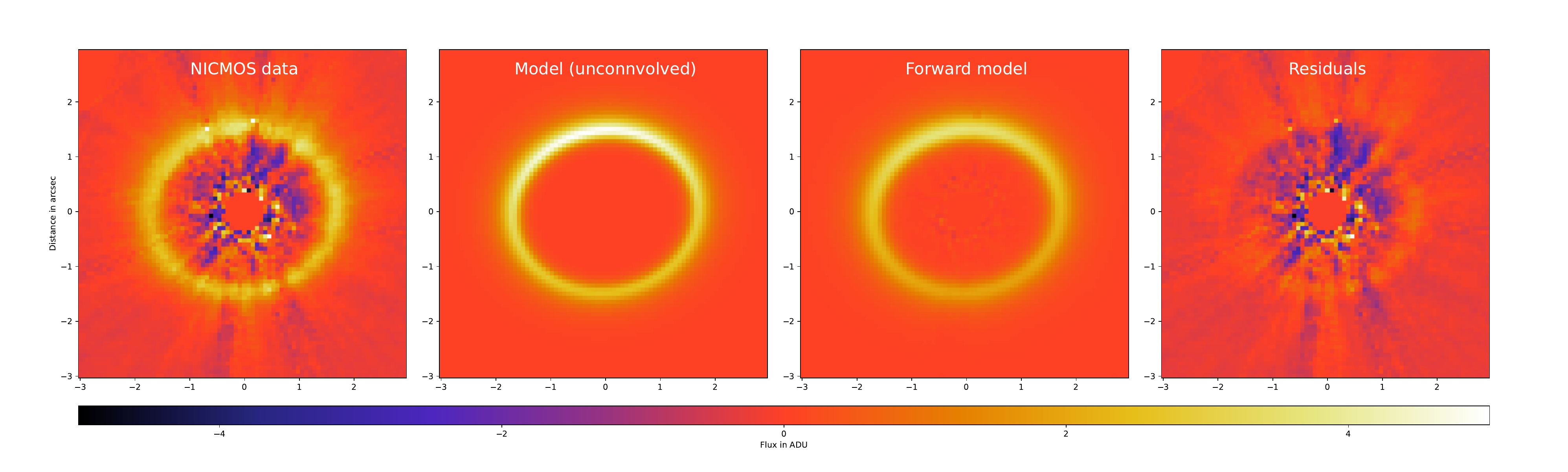}
   \caption{Interpretation of the NICMOS total intensity data. From left to right: NICMOS image reduced with PCA RDI, best model explaining the data, same model after forward modelling, residuals left after forward modelling, and subtraction from the data.} 
    \label{fig_NICMOS_image_model_residuals}
    \end{figure*}

\subsection{Combining total intensity and polarimetry}

As explained in section \ref{sec_strategy}, we now combine the constraints from the IRDIS $Q_\phi$ image and the NICMOS $I_{NIC}$ image in order to find the best model reproducing the IRDIS intensity image $I_{IRD}$.
We do not expect strong chromatic variations of the SPF in the range of scattering angles $60^\circ-120^\circ$. \citet{Chen2020} found for instance no dependence of the near-infrared SPF between $1.1\micron$ and $2.2\micron$ on the system HR\,4796. We therefore assumed that the SPF derived from NICMOS at $1.104\micron$ is still valid at the IRDIS wavelength of $1.625\micron$, and used the parameters $s_{60,I}$, $s_{90,I}$, $s_{120,I}$ in the model reproducing the disc seen in IRDIS total intensity. We refer the reader to the Appendix \ref{subapp_assumption_validation} for a discussion on the validity of this assumption using the dust model introduced later in Sect. \ref{sec_Si_C_ice}. As these parameters also encode the overall brightness of the disc, we need to introduce a scaling factor $\Lambda$ to match the surface brightness of disc model in the  $I_{IRD}$ image. We re-used the parameters $\alpha_{out,pI}$ and $r_{0,pI}$ because we do not expect the morphological parameters to vary between the polarised and total intensity scattered light of the disc. 
Similarly as for NICMOS, we estimated the quality of the model using a chi squared $\chi_{I,IRD}^2$ computed in a circular annulus where disc signal is detected with a specific mask avoiding the artefacts in the image:
\begin{equation}
\chi_{I,IRD}^2 = \displaystyle\sum_{i}    \left[ \frac{I_{IRD}-FM_{I,IRD}(s_{60,I},s_{90,I},s_{120,I},\alpha_{out,pI},r_{0,pI},\Lambda)}{\sigma_{I,IRD}} \right]^2,
\end{equation}
where $\sigma_{I,IRD}$ is the noise map associated to the $I_{IRD}$ image and $FM_{I,IRD}$ is the forward model obtained after convolution and processing of a disc model by the RDI reduction algorithm.

In order to propagate correctly the error bars derived from each individual fit, we performed a global optimisation of the sum of the individual chi squared $ \chi^2_{sum} = \chi_{pI}^2 + \chi_{I,NIC}^2 + \chi_{I,IRD}^2 $. We maximised the log likelihood defined as $-\frac{1}{2}\chi^2_{sum}$ in an MCMC framework, using the same uniform priors as those described for the individual optimisation of the IRDIS polarised intensity model and the NICMOS total intensity model, on top of which we added a uniform prior on the scaling factor $\Lambda$. The results are summarised in the last column of Table \ref{tab_results_mcmc}. The results are compatible within error bar with those obtained on the individual minimisation. As this global optimisation is the only way to obtain robust error bars on our fitted parameters, including the polarised and total intensity phase function, we use these results later on in the paper.

\begin{figure*}
    \centering
   \includegraphics[width=\hsize]{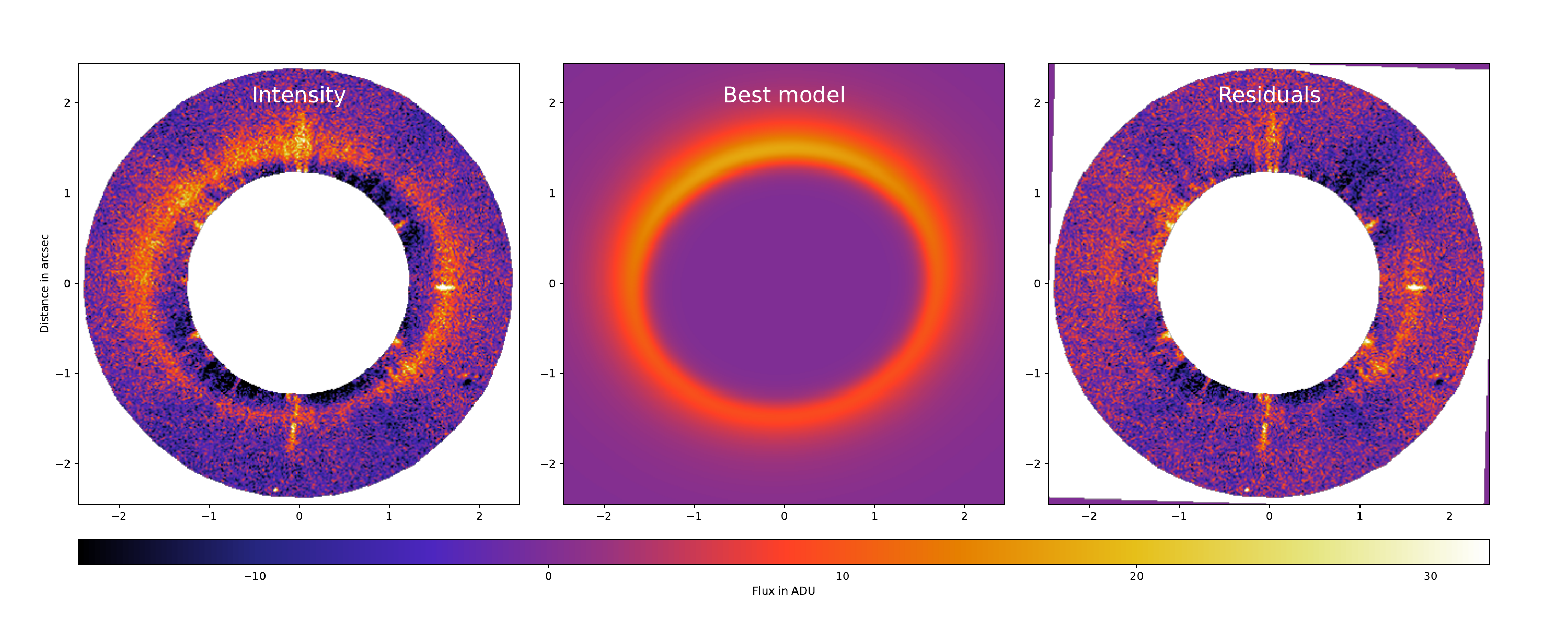}
   \caption{Interpretation of the IRDIS total intensity data. Left: IRDIS total intensity image. Middle: best unconvolved model from the combined MCMC fit. Right: residuals after forward modelling and subtraction of the best model (right).} 
    \label{fig_SPHERE_intensity_image_model_residuals}
    \end{figure*}

\begin{figure}
    \centering
   \includegraphics[width=\hsize]{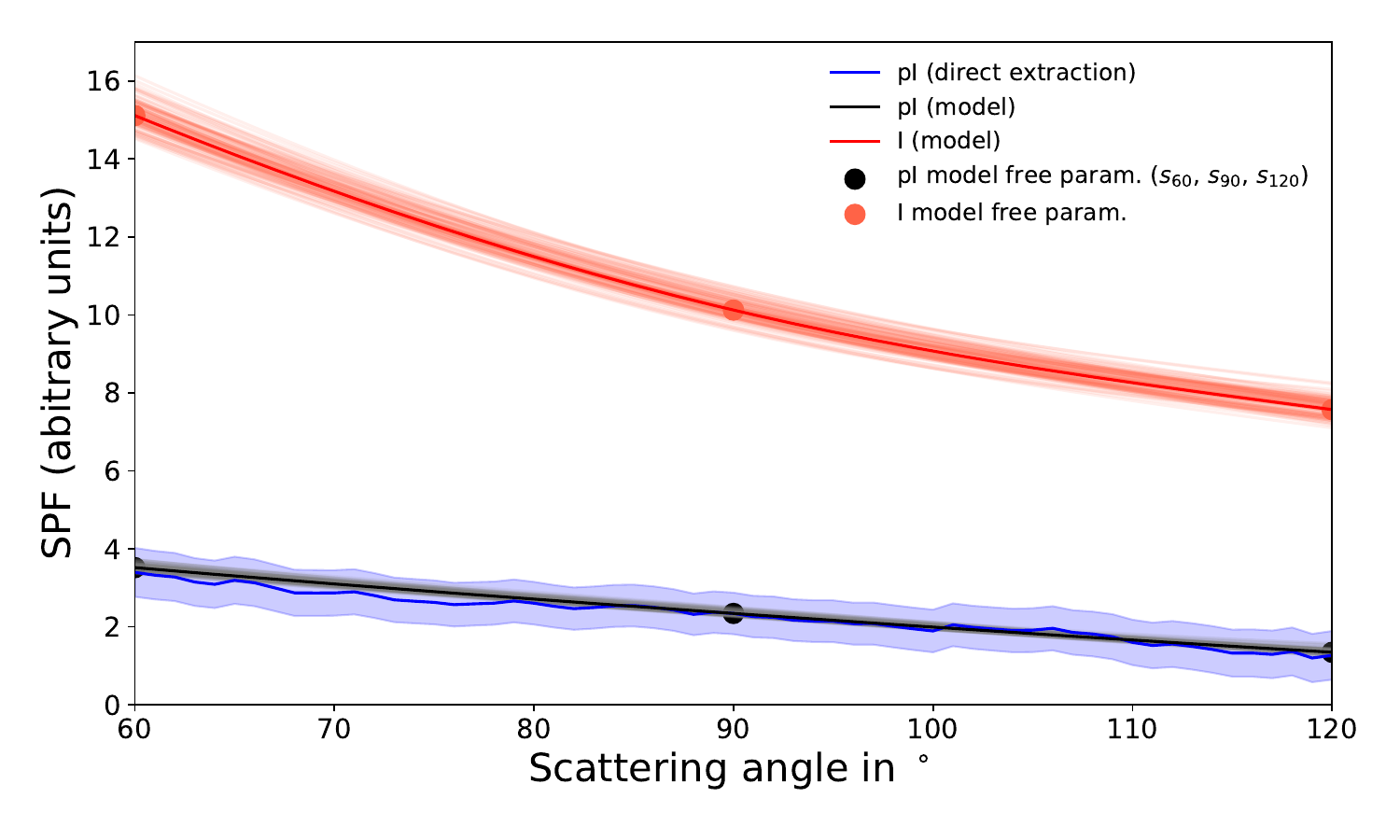}
   \caption{SPF and pSPF on the same vertical scale (in arbitrary units) as a function of the scattering angle} 
    \label{fig_SPF_pSPF}
    \end{figure}

We show in Fig. \ref{fig_SPHERE_intensity_image_model_residuals} the best IRDIS total intensity model (middle planel), and the residuals (right panel) after computation and subtraction of the forward model. Thanks to this combined fit, the pSFP and SPF can be shown on the same scale in Fig. \ref{fig_SPF_pSPF}. The disc in polarised light is about five times fainter than in total intensity.

\subsection{Polarised fraction}
\label{sec_polar_frac}

\begin{figure}
    \centering
   \includegraphics[width=\hsize]{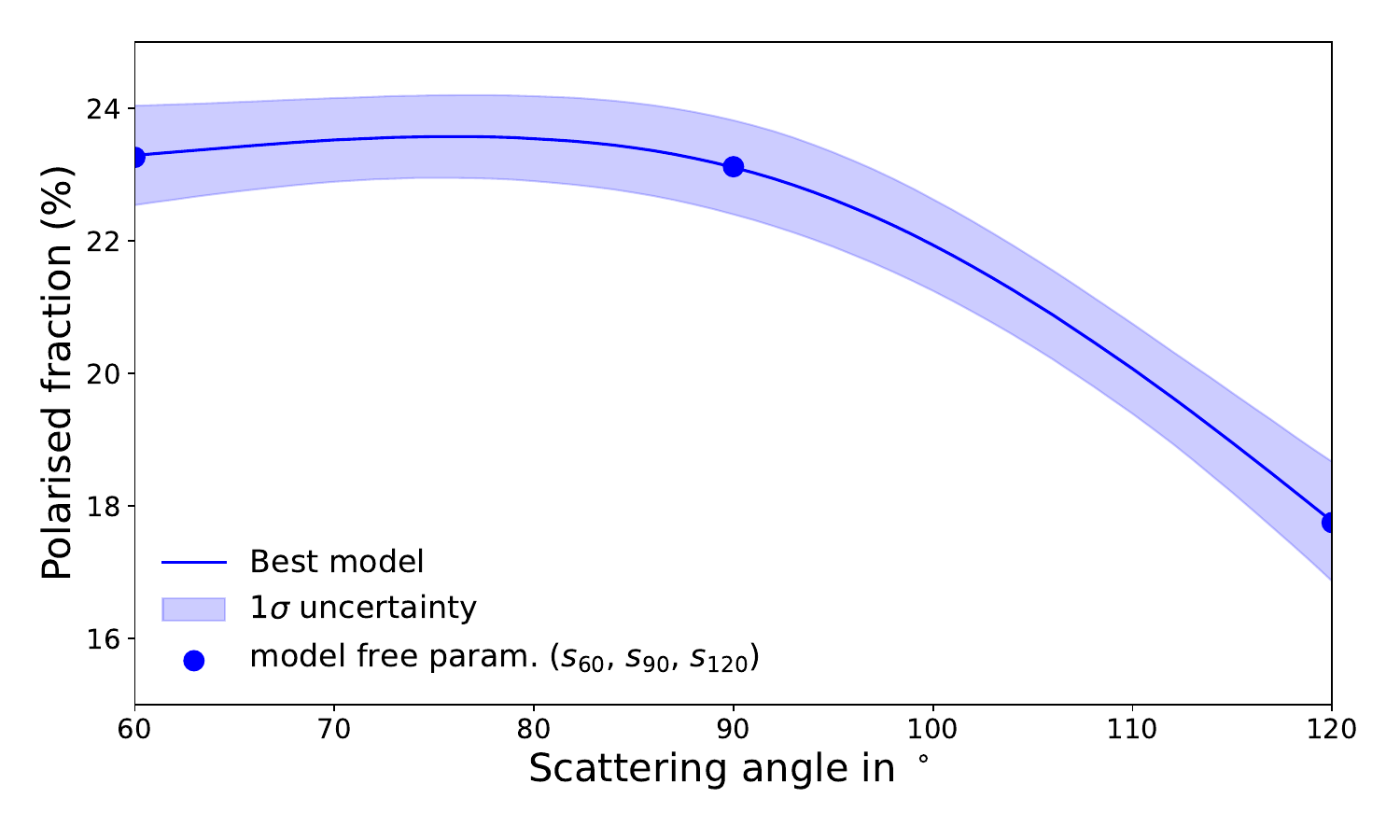}
   \caption{Polarised fraction as a function of the scattering angle} 
    \label{fig_p_frac}
    \end{figure}

The polarised fraction is the ratio of the pSPF by the SPF. It is shown in Fig. \ref{fig_p_frac} after propagating the error bars from the MCMC. The polarised fraction is maximum at about $23.6\% \pm 2.6\%$. The position of the maximum is not well constrained and occurs between scattering angles of $70^\circ$ and  $82^\circ$. The polarisation curve is asymmetric: it is decreasing beyond $100^\circ$ and flatter below $80^\circ$.

An additional result from this modelling is the total scattered intensity of the disc relative to the star flux, $F_{\text{disc},\lambda}/F_{\star,\lambda}$, expressed in the H band. Using the best model presented in Fig. \ref{fig_SPHERE_intensity_image_model_residuals} and the stellar flux estimation presented in section \ref{sec_pdi_technique}, we obtain  $F_{\text{sca},\lambda,\text{disc}}/F_{\star,\lambda}=0.082\% \pm 0.004\%$. This value is twice smaller than that found in the optical with STIS \citep{Schneider2014}, which is  consistent with the fact that the disc is seen as blue between STIS and NICMOS, with $\Delta\text{mag}(\text{STIS}-\text{F110W})=-0.9$ \citep{Ren2023}, equivalent to a flux ratio of $\sim2.3$. The same quantity can be computed for the total polarised scattered intensity of the disc, $(F_{\text{sca},\lambda,\text{disc}})_\text{pol}/F_\star$, using the best model presented in Fig. \ref{fig_polar_disc_model}. We derive $(F_{\text{sca},\lambda,\text{disc}})_\text{pol}/F_{\star,\lambda}=0.019\% \pm 0.001\%$. The uncertainty assumes a $5\%$ accuracy on the estimation of the stellar flux. These numbers can be compared to the infrared excess of the system of $0.2\%$ \citep{Lebreton2012}, to estimate an effective scattering albedo $\omega_{\lambda,\text{disc}}$ defined as the relative contribution of scattering by the disc into the line of sight to the total amount of stellar flux attenuation by scattering and absorption \citep{Engler2023}. Using their Eq. 9, this yields  
$\omega_{\lambda,\text{disc}} = \frac{F_{\text{sca},\lambda,\text{disc}}/F_{\star,\lambda}}{F_{\text{sca},\lambda,\text{disc}}/F_{\star,\lambda}+L_\text{IR,disc}/L_\star} = 0.29 \pm 0.01$.

\section{Interpretation and discussion}
\label{sec_modelling}

We discuss in this section the interpretation of the derived scattering properties with various scattering theories and experimental measurements or observations.

\subsection{Interpretation with spherical particles made of amorphous carbon, silicates and water ice}
\label{sec_Si_C_ice}

\citet{Lebreton2012} showed that the SED of the disc can be well reproduced assuming spherical dust particles made of a mixture of silicates (12\%), carbonaceous material (23\%), amorphous ice (65\%), with 65\% porosity. Their best fit model is found for particles larger than $a_{{min}} = 0.81\micron \pm 0.31 \micron$, with a differential power-law size distribution of exponent $\nu=-3.41 \pm 0.09$. In a first step we therefore investigated the compatibility of this type of dust population with the scattered light properties extracted from our new observations. 
%The best fit model of \citet{Lebreton2012} is purely based on the SED and yields a mass of dust smaller than 1\,mm of $0.051 \pm 0.016 M_\oplus$ incompatible with the surface 
In \citet{Milli2015}, \citet{Milli2019} or \citet{Singh2021}, we computed the theoretical SPF and polarised fraction for a sample of 6500 dust compositions and sizes, using the Mie theory and the distribution of hollow spheres \cite[DHS,][]{Min2005} as provided in the radiative transfer code MCFOST \citep{Pinte2006}.  We reused these models to investigate their compatibility with the observations of \HD. These models are based on a porous dust particle composed of a mixture of astronomical amorphous silicates \citep{Draine1984}, carbonaceous refractory material \citep{Rouleau1991}, and water ice \citep{Li1998} partially filling the holes created by porosity. However, in this work we find that none of these models could satisfactorily reproduce both the total intensity SPF and polarised fraction of the dust particles. 
We refer to Appendix \ref{app_si_ac_ice} for the details on the modelling approach and results. We note that a tension between the scattering properties and the SED was already reported in \citet{Lebreton2012} and \citet{Schneider2006}: the best SED model predicts an albedo 4.5 times larger than what can be inferred from the NICMOS scattered light image, and the total intensity SPF is only weakly forward-scattering while the models predict for particles about the size of the wavelength much higher degrees of forward-scattering. Here, we reveal a new tension when taking into account the polarised fraction. 
We therefore look for solutions by relaxing one of the two assumptions made initially: we kept a similar composition but assumed a different shape for the particles (section \ref{sec_zubko}) or we kept the spherical shape assumption but explored a wider range of optical indices (section \ref{sec_optical_index_search}).

\subsection{Interpretation with a model of agglomerated debris particles}
\label{sec_zubko}

In an attempt to find more compatible models for compact irregular particles, we explored the agglomerated debris particles developed in \citet{Zubko2009}, and used to interpret the scattering properties of cometary or circumplanetary dust \cite[e.g.][]{Zubko2014,Arnold2019,Zubko2020}. They are a model of cosmic dust particles that approximately reproduces the highly irregular morphology of dust in the Solar System. Their morphology is similar to what was detected by Rosetta in the micron-sized dust particles of comet 67P/Churyumov–Gerasimenko \citep{Bentley2016}. This model uses the discrete dipole approximation \cite[DDA,][]{Purcell1973} to compute the scattering properties. 

A population of aggregated debris particles is generated with size  parameters $x$ between 1 and 32, where $x=\frac{2\pi s}{\lambda}$ in a differential power-law size distribution of exponent $\nu$. At $\lambda=1.6\micron$, the particle sizes range therefore between $0.25\micron$ and $8\micron$, but the exact minimum and maximum size does not impact the results for the exponents $\nu$ between $-1.5$ and $-3$ explored here \cite[cf the study in][for details]{Zubko2020_impact_size}. The material is made of a mixture of silicates supposed to be represented by an optical index $(n,k) = (1.6,0.005)$, where $n$ and $k$ represent the real and imaginary part, and amorphous carbon with $(n,k) = (2.43,0.59)$. The agglomerated debris particles have a highly irregular morphology \cite[see][for details and illustrations]{Zubko2020_impact_size} resembling what has been found in cometary dust in situ. We manually optimised the volume fraction of silicates and amorphous carbon to find a model matching both the polarised fraction and total intensity SPF (Fig. \ref{fig_comp_DHS_best_model_combined_observation})

We define a goodness of fit using a reduced chi squared to measure the distance between the model and our measurements, both for the polarised fraction ($\chi^2_\text{pf}$)  and for the total intensity SPF ($\chi^2_\text{SPF}$), and overall ($\chi^2_\text{pf+SPF}=\chi^2_\text{pf} + \chi^2_\text{SPF}$). We note that while the polarised fraction is an absolute value that can be directly compared to that of a model, the model SPF is scaled to match the measured NICMOS SPF in the least square sense before computing the $\chi^2_\text{SPF}$. We found that the model presented in Fig. \ref{fig_comp_DHS_best_model_combined_observation} achieves a good fit for the polarised fraction with $\chi^2_\text{SPF}=2.7$ and is slightly worse for the total intensity SPF with $\chi^2_\text{SPF}=7.6$, for an overall $\chi^2_\text{pf+SPF}=10.3$. The peak polarisation of 23.5\% at a scattering angle of (or below) $80^\circ$ can be well reproduced by the model. The total intensity SPF is more forward-scattering than the actual observations. This difficulty was already mentioned in \citet{Stark2014} with a model based on purely spherical particles. 
We considered alternative highly absorbing component with $(n,k)=(1.855,0.45)$ \citep{Jenniskens1993} or more absorbing silicates with $(n,k) = (1.6,0.01)$ instead of $(1.6,0.005)$, meaning for instance a higher fraction of iron in the generally Mg-rich silicate. However none of the those alternative models helped to improve the fit for the total intensity SPF. %We note however that the overall fit is already a factor 20 better than assuming a spherical shape 
A similar tension was already observed for the disc around HR\,4796 where no Mie or DHS models could simultaneously reproduce the SPF and polarised fraction \citep{Arriaga2020}, despite a thorough grid search through all possible optical indices independently of the underlying physical material making up the dust. Inspired by this study and their grid search, we explored the optical indices of the dust over a much larger range, and kept the assumption of  spherical particles.

\subsection{Interpretation with spherical particles with a parameterised optical index}
\label{sec_optical_index_search}

We generated the scattering properties of spherical particles in a differential power-law size distribution of exponent $\nu$ between $s_{min}$ and 1mm, and with optical indices $(n,k)$. We used the tool \textit{OpTool} \citep{Dominik2021}, implementing both the Mie and DHS theory \citep{Min2005}. The range of parameters explored are summarised in Table \ref{tab_grid_n_k}.

\begin{table}
\caption{Grid of parameters for the 16640 models generated.}             % title of Table
\label{tab_grid_n_k}      % is used to refer this table in the text
\centering                          % used for centering table
\begin{tabular}{p{1.7cm} p{1cm} p{1cm} p{1cm} p{1.3cm}}        % centered columns (4 columns)
\hline\hline                 % inserts double horizontal lines
Parameter & Min. value & Max. value & $N_\text{sample}$ & Sampling  \\    % table heading 
\hline    
Theory &  \multicolumn{2}{c}{Mie / DHS} & / & / \\   % inserts single horizontal line
$s_{min}$ (\micron) & 0.1 & 100 & 13 & log. \\      % inserting body of the table
$n$ & 1 & 4 & 16 & linear \\      % inserting body of the table
$k$ & 1e-7 & 1e2 & 10 & log. \\      % inserting body of the table
$\nu$ & -2.5 & -5.5 & 4 & linear \\
\hline                                   %inserts single line
\end{tabular}
\end{table}

\begin{figure*}
    \centering
   \includegraphics[width=\hsize]{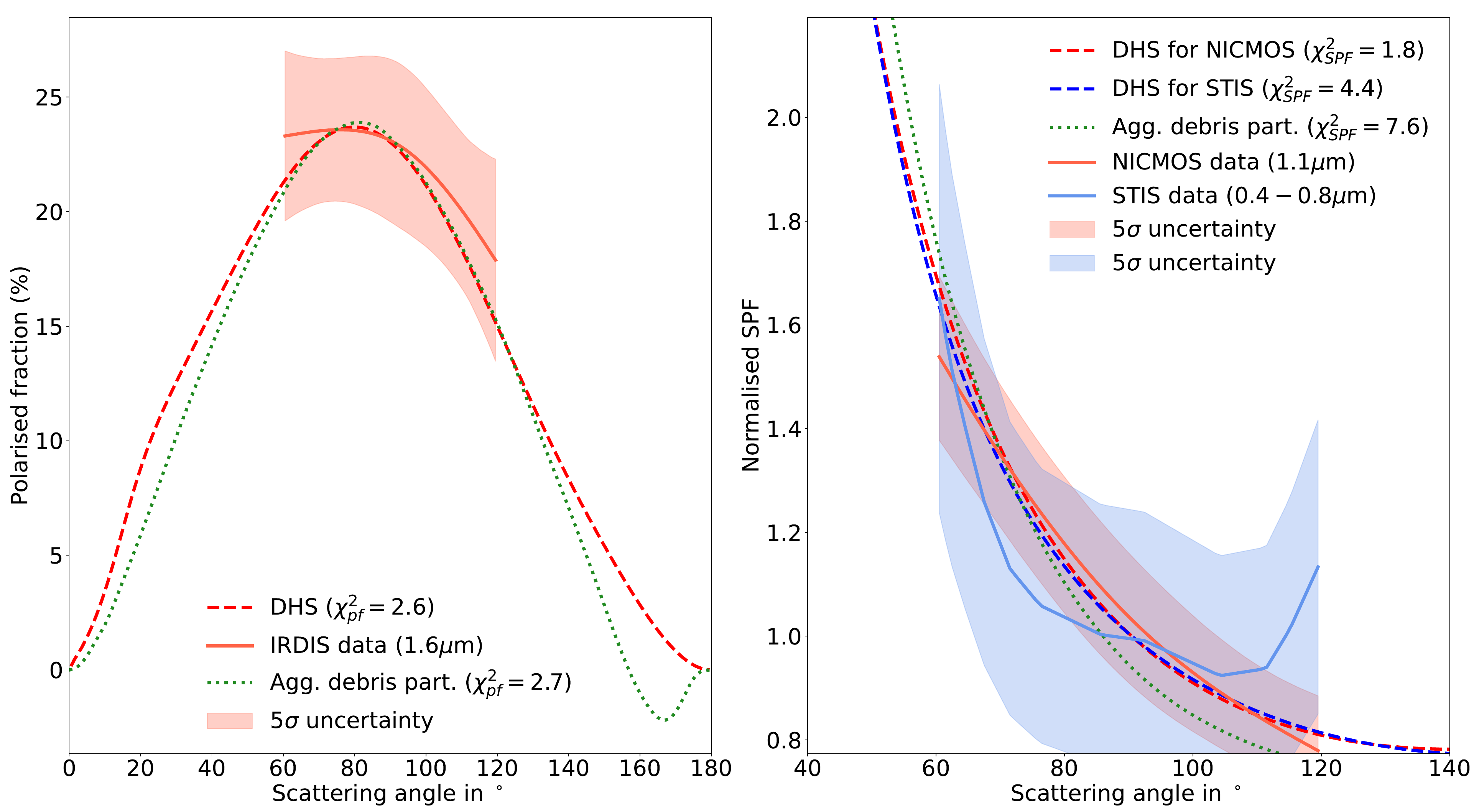}
   \caption{Comparison of the best model among the grid detailed in Table \ref{tab_grid_n_k} (dashed line) with the IRDIS polarised fraction (left, plain red line) and the NICMOS total intensity SPF (right, plain red line). This model provides the smallest reduced $\chi^2_\text{pf+SPF}$ of 4.4.
   The STIS total intensity SPF from \citet{Stark2014} is also shown (right, plain blue line) and compared with the model prediction at the STIS wavelengths (dashed blue line). 
   The two panels also show in green dotted lines one model of aggregated debris particle from \citet{Zubko2009}, matching well the polarised fraction but slightly less the total intensity SPF. This model employs 40\% of the transparent material $1.6+0.001i$ and 60\% of more absorbing material $2.43+0.59i$ in a size distribution parameterised with $\nu=-2.3$, and provides a reduced $\chi^2_\text{pf+SPF}$ of 10.3.  } 
    \label{fig_comp_DHS_best_model_combined_observation}
    \end{figure*}

While the Mie theory failed to give an appropriate model with an overall  $\chi^2_\text{pf+SPF}$ below 15 (i.e. better than the aggregated debris particle shown before), the DHS theory yields four models below this threshold, the best one having $\chi^2_\text{pf+SPF}=4.4$. It is shown in Fig. \ref{fig_comp_DHS_best_model_combined_observation}.

\begin{figure*}
    \centering
   \includegraphics[width=\hsize]{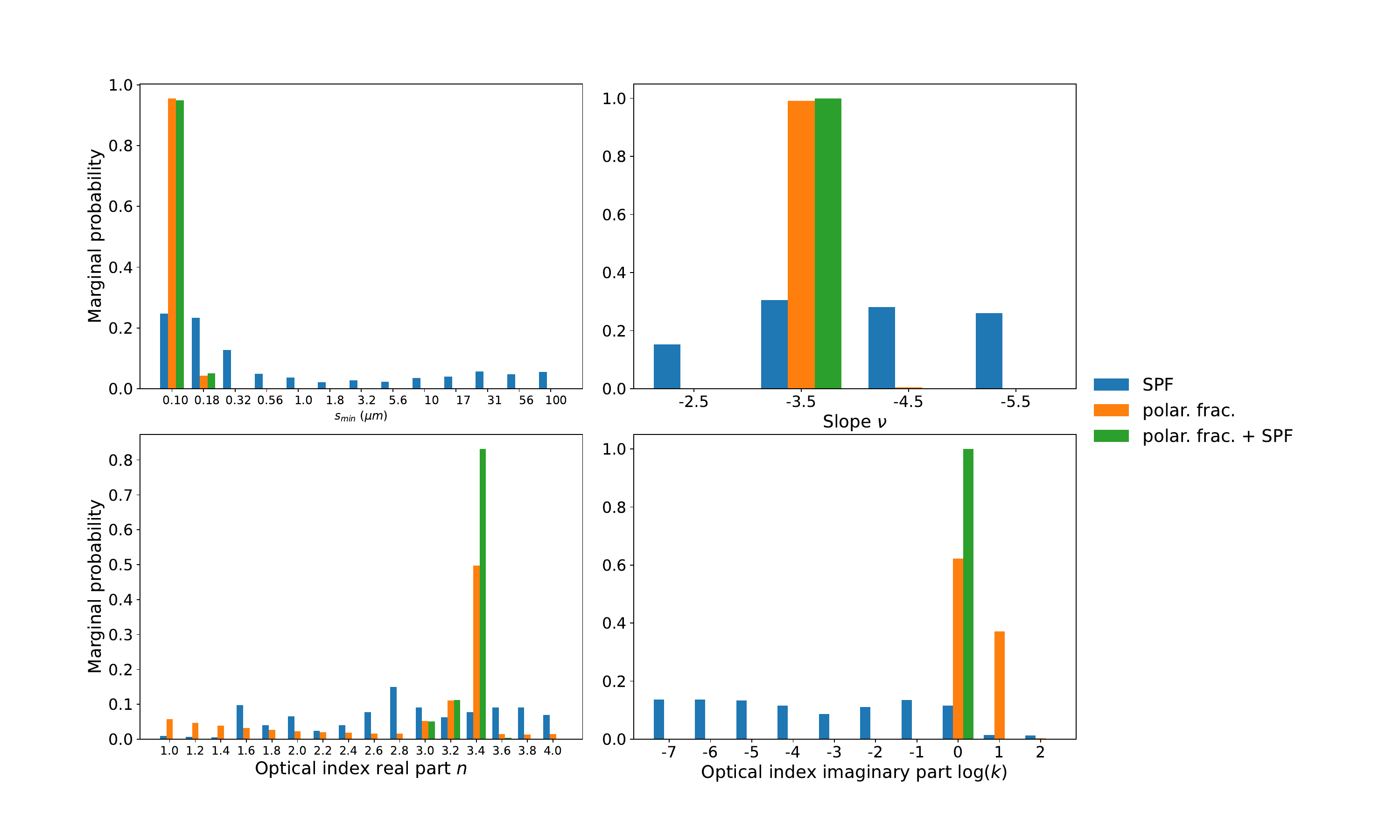}
   \caption{Marginal probability distributions of the four free parameters of our models, based on the polarised fraction (orange bars), on the NICMOS SPF (blue bars) or on the combination of both (green bars). These distributions are derived from models created using the DHS theory} 
    \label{fig_marginal_pdf_DHS}
    \end{figure*}

To provide an estimate of the range of acceptable models for each goodness of fit estimators, we carry out a Bayesian analysis \cite[e.g.][]{Pinte2006,Duchene2010} and assign a probability that the data are drawn from the model parameters. We do not have any a priori information on these parameters, we therefore assumed a uniform prior, corresponding to a uniform sampling of our parameters by our grid (see Table \ref{tab_grid_n_k}). The probability that the data corresponds to a given parameter set is given by $\Psi=\Psi_0\text{exp}\left( -\frac{\chi^2}{2}\right)$ where $\Psi_0$ is a normalisation constant introduced so that the sum of probabilities over all models is unity. The probability given here is only valid within the framework of our modelling and parameter space. Fig. \ref{fig_marginal_pdf_DHS} shows the inferred probability distribution for each of our four free parameters, after marginalisation against all other three parameters. It is shown here using the DHS theory because the best models are obtained with this theory. 

While the total intensity SPF taken alone is not very constraining, the polarised fraction is, and when combined together those two observables show that a power-law exponent of $-3.5$ is very much favoured by the data. This is consistent with the value predicted by the theoretical collisional cascade in the birth ring \citep{Dohnanyi1969}, although theoretical models of size distributions show very complex distributions near the blow-out size \citep{Thebault2019}.  Sub-micron particles are favoured with $s_{min}<0.2 \micron$, together with a relatively large imaginary part $k$ in the range 1-10, meaning highly absorbing material. 

\subsection{Discussion on the size}
\label{sec_discussion_size}

Assuming a spherical shape for the particles, the $0.1-0.2 \micron$ minimum particle size is a robust conclusion of our analysis. A sub-micron minimum size of $0.81 \pm 0.31 \micron$ was also favoured by the SED analysis of \citet{Lebreton2012}, in a power-law distribution consistent with the theoretical collisional cascade. \citet{Stark2014} had already noticed that the total intensity SPF extracted from optical observations was compatible with the Mie theory assuming $\sim0.1 \micron$ particles but noted that such small grains are below the blow-out size and would not survive long enough in the system to dominate the scattering cross section over the bound grains. Our conclusion here is even more robust, because we combine the SPF with the polarised fraction which also strongly favours sub-micron particles. It could also explain why the disc colour is blue between the optical and the near-infrared \citep{Ren2023}.
The blow-out size is estimated to be $\sim 5\micron$, based on the best Mie model of \citet{Lebreton2012} assuming a 65\% porous ice-rich composition, that is 50 times larger than the particles inferred here. Even if ones assumes compact particles with different compositions, the smallest blow-out size could be as low as $0.8 \micron$ \citep{Arnold2019} which is still five times larger than what is inferred here, and no realistic compositions yield a $\beta$ parameter lower than the blowout threshold of 0.5 for particle radii above $0.05 \micron$ \cite[see Fig. 6 b in][]{Arnold2019}. Such small sub-micron grains should thus be blown away from the system on dynamical timescales and their presence might thus appear counter-intuitive.
%The preference for sub-micron particles may have two origins. 
%First, this may be real, meaning that there is 

The possibility that a population of sub-blowout particles leaves a detectable imprint in scattered light and in the thermal SED of the dust is a scenario that was already proposed in \citet{Lebreton2012} and \citet{Stark2014}, or even earlier for debris discs such as HD\,141569A \citep{Augereau2004}. This scenario gained credence when the numerical and theoretical study of \citet{Thebault2019} showed that, for debris discs at collisional steady-state, there is in fact always a non-negligible fraction of unbound grains present in the system. This is especially the case for bright debris discs with infrared excess higher than 0.1\%, for which the population of sub-blowout particles always leaves a detectable signature in scattered light. The contribution of unbound particles with respect to the overall disc flux at $1.6 \micron$ reaches 40\% in their simulation of an A6V star with a very bright disc of infrared excess 0.5\% (see their Fig. 4.b). In the thermal regime, this contribution can even be higher, reaching 80-90\% at wavelengths in the narrow 10 to $20 \micron$ domain, which may very well explain the result of \citet{Lebreton2012}. 
There is, however, a limitation to these simulations, which is that the scattering phase function was assumed to be isotropic. Particles larger than the wavelength are indeed more forward-scattering than sub-micron particles, and for low-inclination discs such as \HD, most of the peak of forward-scattering is not visible, which tends to limit the contribution of those larger particles in the flux scattered towards an observer \citep{Mulders2013}. In a recent study, \cite{Thebault2023} explored these issues and found that, for a belt+halo system such as \HD{}, realistic scattering phase functions generally tend to increase the contribution of unbound grains to the flux in scattered light. This contribution can reach 10 to 30\% in the parent ring and up 50 to 70\% in the extended halo (see Fig.2 of that paper). In addition, the presence of gas may also increase the fraction of sub-blowout particles by maintaining them longer than expected in the main parent belt because gas friction might slow down, or even halt their outward motion \citep{Takeuchi2001}. This scenario was proposed for HD\,32297 \citep{Bhowmik2019}.
%Therefore, the signature of sub-blowout particles in the system may be naturally explained by the high infrared excess and luminosity of the star. 

Another possibility is to invoke a recent catastrophic collision, which would produce a population of particles outside collisional equilibrium, with a large fraction of unbound grains \citep{Kral2015}. Such a massive collisional event was proposed in \citet{Stark2014} to explain a density enhancement seen in the disc at optical wavelength, but we do not detect such an asymmetry in the near-infrared. We note also that the large quantities of unbound grains produced by such a transient catastrophic event would be blown out of the system on relatively short timescales \citep{Jackson2014}, thus making the probability to witness such an event very low.

We can, however, not rule out the fact that the preference for sub-micron particles might be an artefact from the DHS or Mie theory, assuming perfectly spherical particles, while real disc particles are likely irregular and rough at their surface. If those irregularities or this surface roughness is of the order of  $0.1 \micron$, DHS may lead to a better fit with $s_{min}=0.1 \micron$ particles, as discussed in \citet{Min2005}.

\subsection{Discussion on the composition and shape of the particles and comparison to Solar-System comets}

\begin{figure}
    \centering
   \includegraphics[width=\hsize]{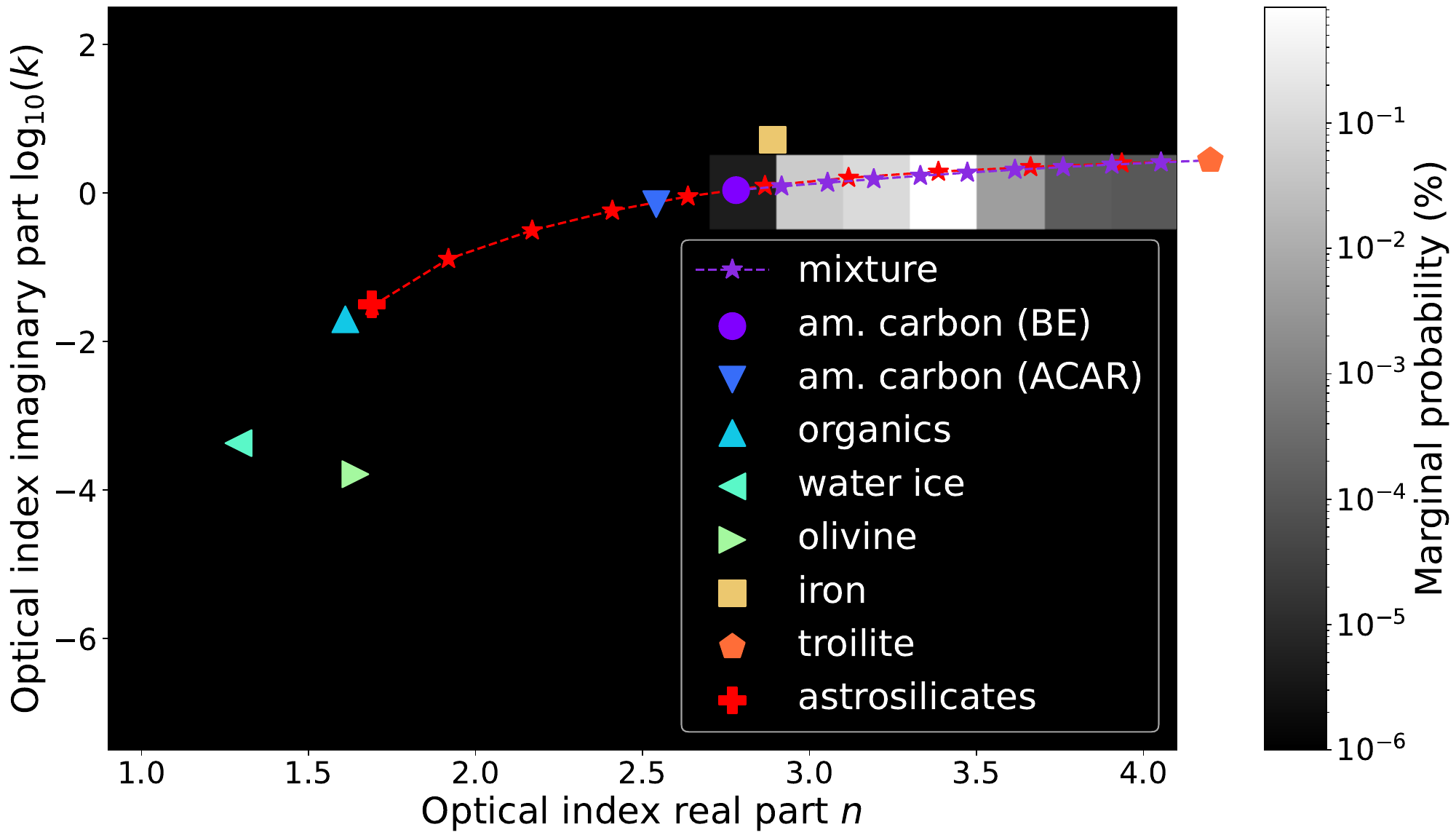}
   \caption{Marginal probability map of the optical index $(n,k)$ of the dust based on the polarised fraction and NICMOS SPF. We overplotted optical constants of amorphous carbon \cite[][BE = BEnzene burning, ACAR = arc discharge between Amorphous Carbon electrodes in an ARgon atmosphere]{Zubko1996}, water ice, olivine organics, iron and troilite from \citep{Pollack1994} and astrosilicates from \citet{Draine1984}. The red (respectively purple) line shows a mixture of astrosilicates (resp. amorphous carbon) with troilite.} 
    \label{fig_joint_marginal_pdf_DHS_n_k}
    \end{figure}

% Zubko+1996: 1.6micron corresponds to E=0.77eV
%tholins at 1.6 mic: $1.65 + 4.05e-4 * i$ \citet{Khare1984}
%From \citet{Pollack1994}:
%water ice $1.29 + 4.3e-4 * i$
%olivine $1.64 + 1.64e-4 * i$
%orthopyroxene: $1.63 + 3.0e-4 * i$
%organics: $1.61 + 2.0e-2 * i$
%iron $2.89 + 5.2 * i$
%troilite $4.2 + 2.77 *i $
%astrosilicates from \citet{Draine1984} $1.6799E+00   3.1822E-02$
%H2O-dominated “dirty” ice (mostly crystalline) compiled from several sources %described in \citet{Li1997}: $1.3136E+00 +  5.4791E-04 * i$
%(amorphous) carbonaceous dust \citet{Rouleau1991}: $2.1088E+00   2.3265E-01$

Our exploration of the optical index of the particles favours a relatively absorbing material ($k\sim1$) and highly refractive ($n\sim3.4)$. In Fig. \ref{fig_joint_marginal_pdf_DHS_n_k}, we show a probability map of the optical index $(n,k)$ based on the combination of the polarised fraction and SPF. It was obtained by marginalising (i.e. integrating) our probability distribution over the minimum particle size $a_{{min}}$ and the slope $\nu$. The sum of the probability of this map is 1, and the map sharply peaks at 83\% for $(n,k)=(3.4,1)$. We see that most standard material used to interpret dust in protoplanetary or debris discs are less refractive ($n \leq 3$), meaning that even a mixture of those would not yield a resultant material refractive enough. Troilite (FeS) is an interesting exception as the real part $n$ of the optical index is as large as 4.2 \citep{Pollack1994} or even 6.8 using the optical constants from \citet{Henning1996}, not represented in Fig. \ref{fig_joint_marginal_pdf_DHS_n_k} as it is beyond the exploration range. There is solid evidence for the widespread occurrence of sulfide minerals, including troilite in primitive bodies of the Solar System. In Antarctic micro-meteorites and in the grains of the 81P/Wild\,2 comet returned by the Stardust mission, the presence of Fe-Ni sulfides is abundant and dominated by troilite \citep{Dobrica2009}. Fe-Ni sulfides are also very good dark and opaque mineral candidate to explain the low reflectance of the comet 67P/Churyumov–Gerasimenko \citep{Quirico2016}. Its presence in a small proportion could explain the large value of $n$ favoured by our model. For a reference, the averaged atomic mass fraction of Fe is 7.5\% in the comet 67P \citep{Bardyn2017} and the main carrier of Fe is supposed to be in the mineral phase, in the form of anhydrous sulfides and Fe- Ni alloys \citep{Quirico2016}. Such a proportion of opaque minerals, including sulfides such as pyrrhotite ($Fe_{{1-x}}S$) and troilite, may possibly also be present in the \HD{} dust particles. 

Interestingly, we also notice on Fig. \ref{fig_joint_marginal_pdf_DHS_n_k} that the best composition lies relatively close to the amorphous carbon optical constants from \citet{Zubko1996}, especially the sample labelled 'BE' with $(n,k)=(2.78,1.097)$, which is widely used to interpret circumstellar dust properties \cite[e.g.][]{Tazaki2023}. The best model of \citet{Lebreton2012} used the amorphous carbon optical constants labelled ACAR, slightly further away from the region favoured by the dust scattered light properties, with a volume fraction of 23\% for this component. We may therefore conclude that the amorphous carbon is a promising candidate if combined with a more highly refractive material such as troilite or other type of dark absorbing mineral. In the $(n,k)$ parameter space represented in Fig. \ref{fig_joint_marginal_pdf_DHS_n_k}, we added the track where a compact (non-porous) mixture of amorphous carbon and troilite would lie (purple line, each symbol along the line representing an increment of 10\% in one of the end members). The Bruggeman mixing rule was used to compute the optical constant of the resulting material. The best material candidate for the \HD{} dust particles correspond to a mass ratio of 60\% amorphous carbon and 40\% troilite. This is an illustration of one possible mixture leading to a real material compatible with the scattered light properties extracted in this work. This is not enough to confirm the presence of troilite or estimate its mass abundance for several reasons. First, there is an infinite number of solutions, using for instance, different end members, more than two materials, or adding porosity to the particles. The red track shows the mix of astrosilicates with troilite. In general, when one assumes only two end members mixed together without porosity, the mixture track is a smooth and curved line joining the two pure end members. Second, if the particles are not homogeneous and consist for instance of a core and an external coating, the bulk mass abundance may be different from relative abundance retrieved from the scattering properties. Last, the shape of the particles may be significantly different from perfect spheres, implying deviations from the Mie or DHS theory not investigated in this study. \citet{Munoz2021} showed that the use of the Mie theory may lead to overestimation of the refractive index $n$ and $k$ and underestimation of the particle size when it is used to fit the polarised fraction of particles of the order of the wavelength or larger. In this respect, the investigation of the agglomerated debris particles presented here show that the of an irregular geometry with only amorphous carbon and silicates alone, already provides a relatively good match to the data without requiring high optical indices.  
%On-going laboratory analysis of the light scattered by irregular compact micron-sized iron sulphides particles show that the peak of polarisation fraction may occur at scattering angles below $30^\circ$, depending mostly on the size of the particles (Milli et al. in prep). 

The main constituent of the model proposed in \citet{Lebreton2012} is water ice, with a volume fraction of 65\%. Here, we see in Fig. \ref{fig_joint_marginal_pdf_DHS_n_k} that the optical constant of water ice lies relatively far from the likely parameter space. This was also a conclusion from section \ref{sec_Si_C_ice} and Appendix \ref{app_si_ac_ice}, where no model including water ice could well reproduce the data. The main impact of the large amount of water ice is to produce a very high albedo of 65\% for the dust. \citet{Stark2014} showed that such a high albedo is compatible with the scattered light observations, if one assumes a realistic SPF to account for the flux scattered outside the line of sight towards the Earth. Our current work cannot rule out the presence of water ice, if it is mixed with other more absorbing materials that would dominate the scattering properties. Observations at longer wavelengths, for instance in the $3.0 \micron$ water absorption band, could confirm the presence of water ice, as it is now possible thanks to the sensitivity of the James Webb Space Telescope observations  \cite[see e.g.][]{McClure2023}. 

%\subsection{Comparison to solar-system comets}

\begin{figure}
    \centering
   \includegraphics[width=\hsize]{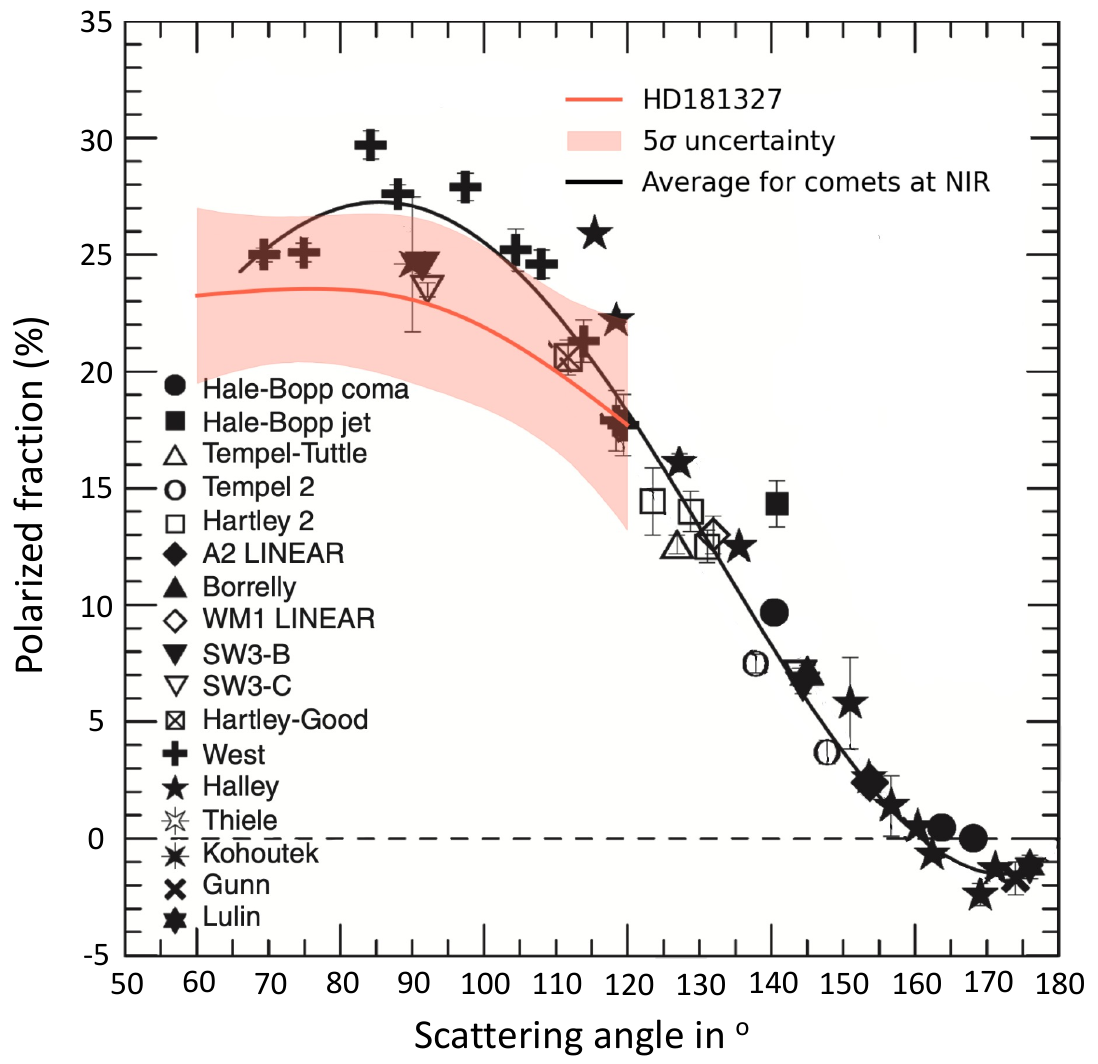}
   \caption{Comparison of the \HD{} polarisation fraction with that from comets in the near-infrared (J, H or K band, adapted from \citep{Kiselev2015}).} 
    \label{fig_kiselev}
\end{figure}

Comets in our Solar System represent an interesting point of comparison because cold debris rings are considered as a reservoir of cometary material releasing smaller particles through collisions \citep[see][for a review]{Kral2018_prospective,Hughes2018}. The dust released in the tail of comets close to their perihelion passage when they sublimate has been widely studied for a number of bright comets, in the optical or in the near-infrared \cite[see][for a review]{Kiselev2015}. We show in Fig. \ref{fig_kiselev} the comparison between the polarisation fraction extracted from the \HD{} particles and that of comets in the J, H or K band. For geometrical reasons, it is easier to observe comets at high scattering angles larger than $120^\circ$, especially beyond $160^\circ$ close to what is referred to as the negative polarisation branch. But the NIR positive branch is now sufficiently documented to allow a meaningful comparison, with an average maximum value $P_{{max}}$ in the 20\%-30\%. The \HD{} dust properties interestingly show a similar behaviour, with $P_{{max}} = 23.6\% \pm 2.6\%$. This is similar to the debris disc HD\,35861 \cite[$P_{{max}} \sim 25\%-30\%$][]{Esposito2018}, but larger than the HD\,114082 disc \cite[$P_{{max}} \sim 17\%$][]{Engler2023} and smaller than HR\,4796 \cite[$P_{{max}} \sim 50\%$][]{Arriaga2020}. 
This analogy with comets can be used to extrapolate extrasolar dust properties from what we know from cometary dust based on in situ or remote observations, but also to guide our modelling approach based on scattering theories already proven successful at modelling cometary dust properties. 
In situ observations of the comet 67P by the Rosetta mission showed that the cometary particles have a variety of morphologies, including compact single elongated grains and larger porous aggregate particles probably formed by the hierarchical agglomeration of these smaller compact grains \citep{Bentley2016}.

Numerical simulations have been developed to reproduce the polarised fraction of some extensively observed comets, such as 1P/Halley, C/1995 O1 Hale–Bopp or 67P/Churyumov-Gerasimenko. Satisfactory fits to polarised observations have been obtained over a wide range of wavelengths with aggregates of submicron-sized grains mixed with spheroidal particles. Both are consisting of absorbing organic-type material and weakly absorbing silicate-type material \citep{Levasseur-Regourd2008,Lasue2009}. With the additional constraint from the total intensity SPF, two other models have proven successful to reproduce cometary dust. First, the rough spheroid model, representing a polydisperse mixture of randomly oriented smooth and rough spheroids of a variety of aspect ratio, could reproduce the intensity and polarisation properties of most cometary dust \citep{Kolokolova2015}. The Gaussian Random Sphere model of \citet{Markkanen2018} could also reproduce the photopolarimetric behaviour of 67P. Those successful numerical models seem to show that the non-spherical nature of the particles seem important to properly model the properties of cometary dust, and their application to circumstellar debris dust is highly relevant now that detailed data have been extracted on a couple of bright debris discs.

%With a maximum polarisation lower than $25\%$, particles made of pure amorphous carbon or pure silicates monomers smaller than $400 \mu m$ and aggregated in larger micron-size clusters are not very likely, as they would produce a higher polarisation fraction at $1.6\mu m$, even for the most compact aggregates \citep{Tazaki2022}.  

%The maximum polarisation of depend on a number of parameters, most importantly the structure of the particles, the minimum particle size and the albedo. If particles are aggregated, the size of the monomer dominates the maximum polarisation, with small monomers compared to the wavelength having a large polarisation fraction \citep{Tazaki2022}. A similar conclusion is reported in \citet{Kiselev2015} for comets: a high polarisation can mostly be attributed to the domination of smaller or more porous particles while smaller polarisation can be attributed to the domination of large compact particles. Last we also highlight the high polarisation fraction by 

\section{Conclusions}
\label{sec_conclusions}

In this paper, we extract the morphology of the \HD{} debris disc, together with its polarised and total intensity scattered light properties, using near-infrared observations with VLT/SPHERE in the H band. We show first that the dust particles scattering predominantly at this wavelength have a different radial distribution from the particles responsible for the thermal emission at $0.88\text{mm}$. Their peak density is offset by 2.3\,au $\pm$ 0.2\,au away from the star. This size segregation can be natural in belts sufficiently wide like \HD, or can result from the low amount of gas detected in this disc in previous ALMA observations. The outer slope is also less steep, as expected from the radiation pressure of the central star bringing smaller particles more sensitive to the radiation pressure on eccentric orbits.

We use forward modelling to extract both the total intensity SPF and the polarised intensity SPF. Due to the challenge at obtaining a high S/N total intensity image to extract the SPF of the dust from the SPHERE data alone, we develop an innovative technique, combining HST/NICMOS to constrain the total intensity SPF with SPHERE to constrain the polarised intensity SPF and morphology, yielding a reliable estimate of the polarised fraction. Such an approach may be used in the future to extract the polarisation fraction of other disks, in order to benefit from the low level of stellar residuals from ground-based high-contrast polarised coronagraphic imaging, and the high sensitivity of space-based total intensity imaging. An interesting prospect is for instance to combine SPHERE/IRDIS polarimetry with JWST/NIRCAM imaging to retrieve the polarised fraction and total intensity SPF of either additional discs or at additional near-infrared wavelengths. 

In the range of scattering angles $60^\circ-120^\circ$ accessible for this disc given the system inclination, the maximum polarisation degree is $23.6\% \pm 2.6\%$. The total intensity SPF is monotonic and smoothly forward scattering, without evidence for backward scattering beyond $90^\circ$, as detected at optical wavelengths. Models based on compact spherical particles made of silicates, amorphous carbon, water ice and porosity that could well reproduce the SED cannot explain simultaneously the polarised and total intensity scattered light properties. We show that sub-micron particles in a power-law size distribution of exponent $-3.5$, made of a highly refractive material and as absorbing as amorphous carbon is required to explain both the polarisation and total intensity SPF in the assumption of spherical particles. Such a material can be obtained, for instance, by mixing iron sulphide with amorphous carbon, but there exists an infinite number of possibilities. 
The dust degree of polarisation is strikingly similar to Solar System comets. We therefore speculate that the dust particles may share similar properties, in particular irregular shapes with a mixture of compact grains and aggregated particles, as shown by the in situ measurements of Rosetta on the comet 67P/Churyumov-Gerasimenko. Future work may therefore look into the application of advanced  scattering models developed to interpret the irregular compact and aggregated cometary dust to circumstellar dust. That could provide more insights into the composition and shape especially if those models can be coupled to radiative transfer to simultaneously reproduce the dust thermal emission and disk SED. 

Access to the optical wavelength range represents another avenue to test the validity of the interpretation in terms of dust composition and size. The optical polarimetric channel of SPHERE, called ZIMPOL \citep{Schmid2018}, can be used to complement NIR polarimetric observations of protoplanetary and bright debris discs \cite[e.g.][]{Milli2019,Ma2023}. The upcoming Nancy Grace Roman Space Telescope Coronagraph instrument \citep{Kasdin2020} is expected to provide a higher sensitivity to faint debris discs and an accurate calibration of the polarisation fraction better than 3\% to allow high-fidelity retrieval of the dust properties in the optical \citep{Anche2023_PASP}.

\begin{acknowledgements}

We dedicate this paper to the memory of Anny-Chantal Levasseur-Regourd (ACLR), who provided significant support and inspiration for that work during the interpretation steps of these observations, as part of the EPOPEE project (Etude des POussi\`eres Plan\'etaires Et Exoplan\'etaires). She contributed significantly to the EPOPEE team, a research group funded by the Programme National de Plan\'etologie (PNP) of CNRS-INSU in France, from which we acknowledge financial support. ACLR was one of the most enthusiastic members, always willing to share her wide expertise with the younger generation, and she has vastly contributed to this scientific collaboration between the disc and Solar System community. She sadly passed away only a few weeks after attending our June 2022 EPOPEE working group meeting in Grenoble. 

SPHERE is an instrument designed and built by a consortium consisting of IPAG (Grenoble, France), MPIA (Heidelberg, Germany), LAM (Marseille, France), LESIA (Paris, France), Laboratoire Lagrange (Nice, France), INAF Osservatorio di Padova (Italy), Observatoire de Gen\`eve (Switzerland), ETH Zurich (Switzerland), NOVA (Netherlands), ONERA (France) and ASTRON (Netherlands) in collaboration with ESO. SPHERE was funded by ESO, with additional contributions from CNRS (France), MPIA (Germany), INAF (Italy), FINES (Switzerland) and NOVA (Netherlands).  SPHERE also received funding from the European Commission Sixth and Seventh Framework Programmes as part of the Optical Infrared Coordination Network for Astronomy (OPTICON) under grant number RII3-Ct-2004-001566 for FP6 (2004-2008), grant number 226604 for FP7 (2009-2012) and grant number 312430 for FP7 (2013-2016). This work has made use of the High Contrast Data Centre, jointly operated by OSUG/IPAG (Grenoble), PYTHEAS/LAM/CeSAM (Marseille), OCA/Lagrange (Nice), Observatoire de Paris/LESIA (Paris), and Observatoire de Lyon/CRAL, and supported by a grant from Labex OSUG@2020 (Investissements d’avenir – ANR10 LABX56).

Finally this project is co-funded by the European Union (ERC) through the projects Dust2Planets n$^\circ$ 101053020, ESCAPE n$^\circ$ 101044152 and COBREX n$^\circ$ 885593. Views and opinions expressed are however those of the authors only and do not necessarily reflect those of the European Union or the European Research Council Executive Agency. Neither the European Union nor the granting authority can be held responsible for them.
The project is also supported by the French National Research Agency (ANR-21-CE31-0015-03) as part of the DDISK project (Disk Discoveries: Investigating morphology and duSt properties with breaKthrough data science).

Ryo Tazaki acknowledges financial support from the CNES fellowship. Alice Zurlo acknowledges support from the ANID -- Millennium Science Initiative programme -- centre Code NCN2021\_080. The work by Evgenij Zubko was supported by the Institute for Basic Science (IBS-R035-C1).

\end{acknowledgements}

\bibliography{biblio}     

%\Online
\begin{appendix} %First online appendix

\section{Model fitting}
\label{App_SPF}

\subsection{Extraction of the SPF from IRDIS data}
\label{App_pSPF_IRDIS}

\begin{figure}
    \centering
   \includegraphics[width=\hsize]{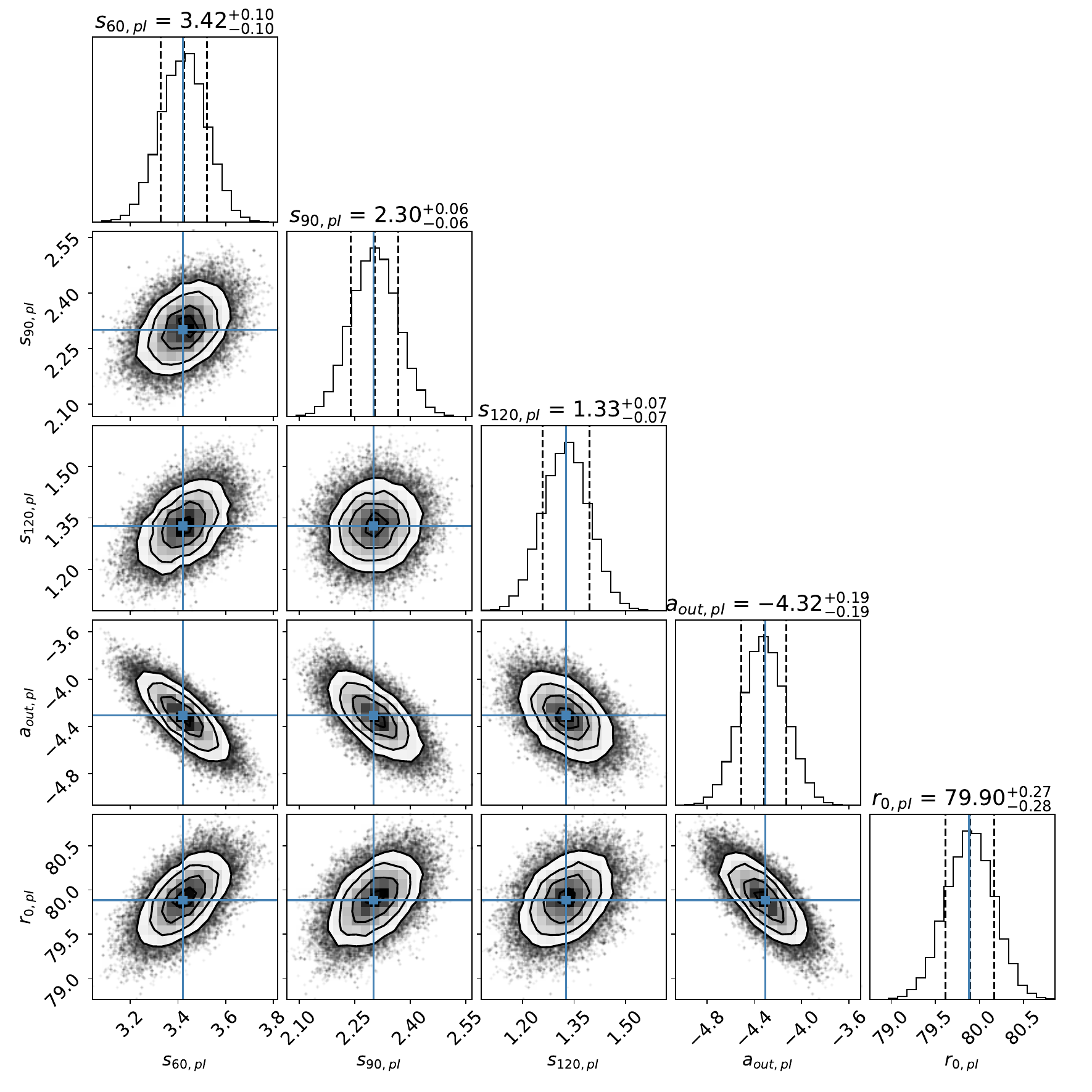}
   \caption{Posterior distribution of the parameters describing the polarised intensity of the disc.} 
    \label{fig_cornerplot_PDI}
    \end{figure}

The posterior probability distribution of the five free parameters. parametrising the polarised intensity disc model constrained by the IRDIS image are shown in Fig. \ref{fig_cornerplot_PDI}.

\subsection{Extraction of the SPF from NICMOS data}
\label{App_SPF_NICMOS}

\begin{figure}
    \centering
   \includegraphics[width=\hsize]{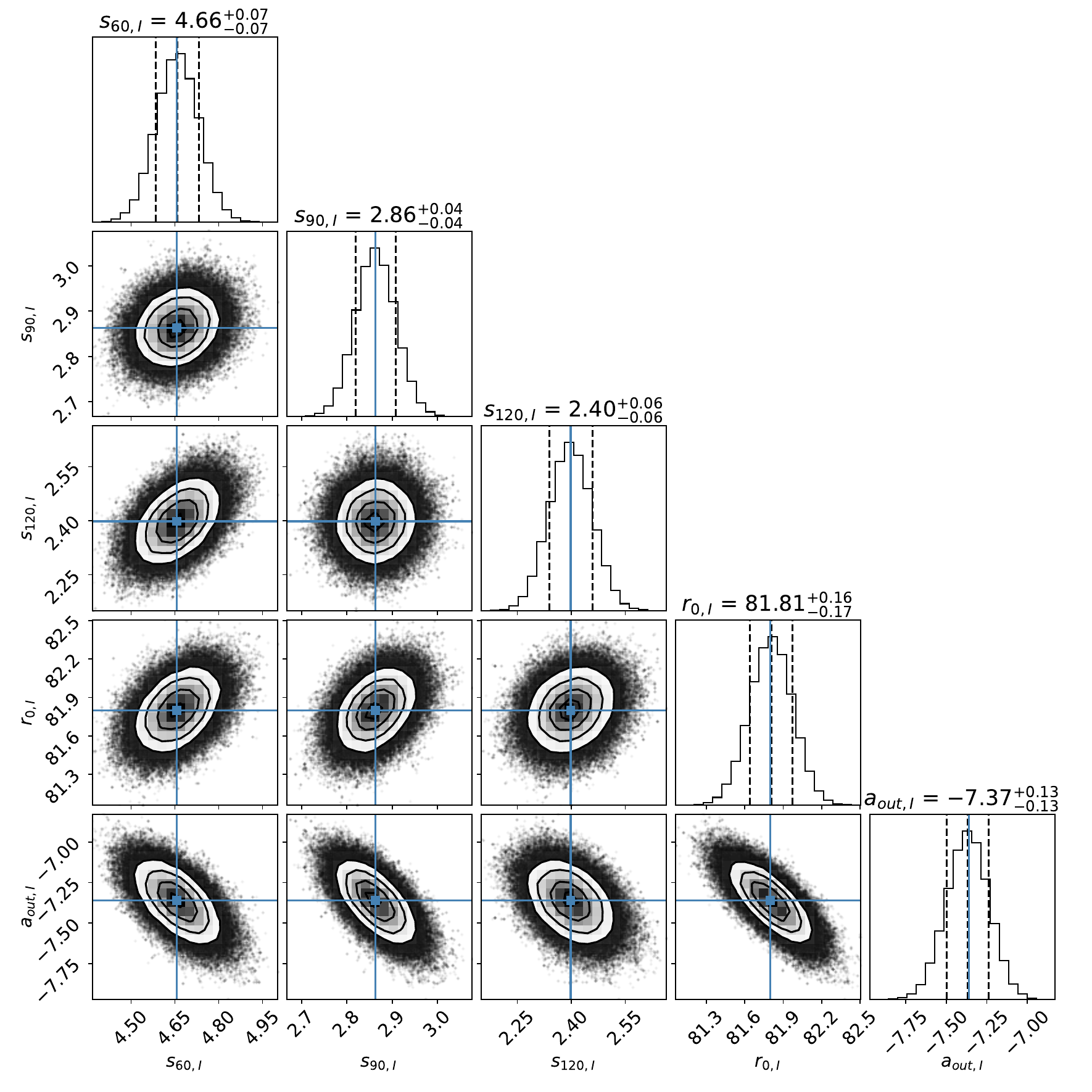}
   \caption{Posterior distribution of the parameters describing the total intensity of the disc with NICMOS.} 
       \label{fig_cornerplot_NICMOS_SPF}
    \end{figure}

The posterior probability distribution of the five free parameters. parametrising the total intensity disc model constrained by the NICMOS image are shown in Fig. \ref{fig_cornerplot_NICMOS_SPF}.

\subsection{Combined extraction of the SPF and pSPF from NICMOS and IRDIS data}
\label{App_SPF_NICMOS_IRDIS}

\begin{figure*}
    \centering
   \includegraphics[width=\hsize]{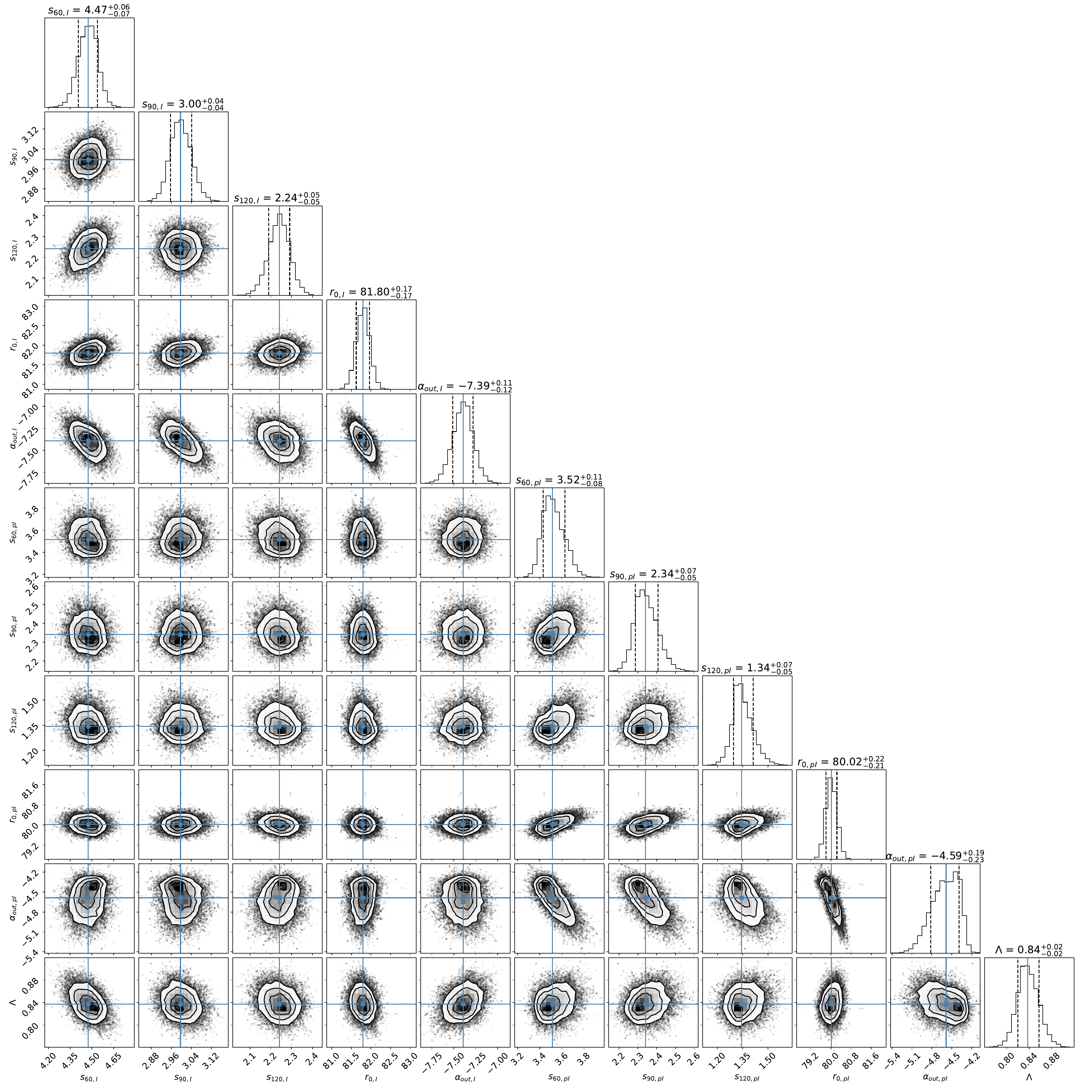}
   \caption{Posterior distribution of the parameters used in the combined fit of the NICMOS and IRDIS total intensity of the disc and the IRDIS polarised intensity.} 
    \label{fig_cornerplot_NICMOS_IRDIS}
    \end{figure*}

The posterior probability distribution of the 11 free parameters. parametrising the total intensity disc model and polarised intensity disc model constrained by the NICMOS and IRDIS image are shown in Fig. \ref{fig_cornerplot_NICMOS_IRDIS}.

\section{Extraction of the total intensity SPF from IRDIS data}
\label{App_SPF_IRDIS}

\begin{figure*}
    \centering
   \includegraphics[width=\hsize]{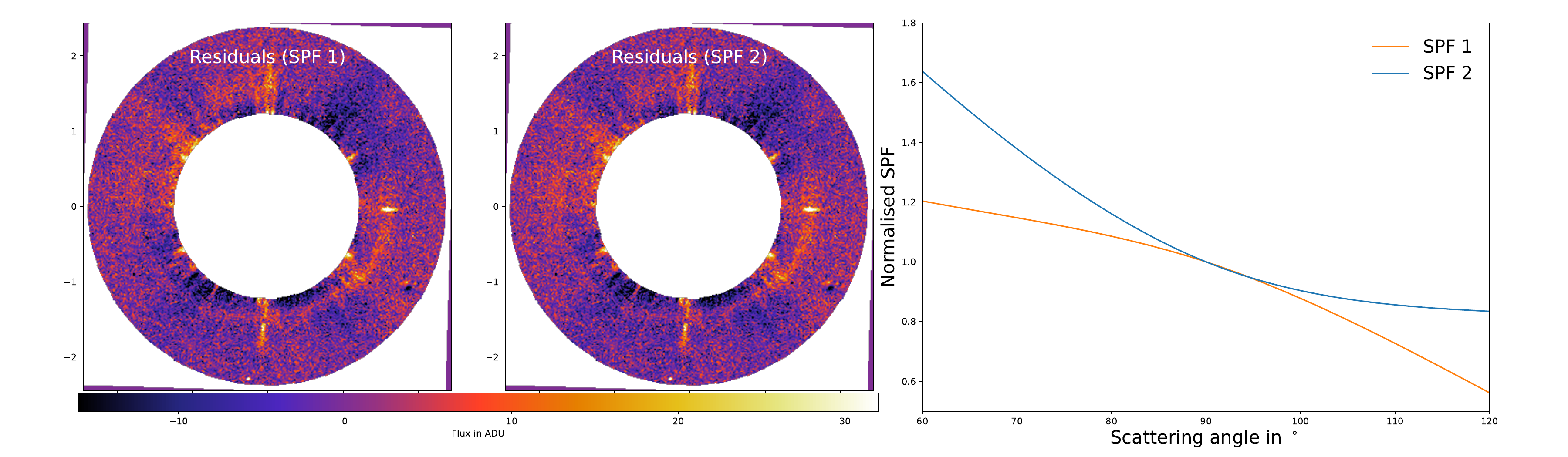}
   \caption{Degeneracy in the SPF extraction from IRDIS alone. The left and middle images show residuals from IRDIS after the subtraction from the data of two disc models using different SPF parametrisations (right graph).} 
    \label{fig_comparison_IRDIS_NICMOS_SPF}
    \end{figure*}

To illustrate the impossibility to retrieve the SPF from IRDIS total intensity data, we show in Fig. \ref{fig_comparison_IRDIS_NICMOS_SPF} the residual image after subtracting from the data two models with different SPF. Both residual images provide a similar chi squared $\chi^2_I$ compatible with the data.
Even though the sharp adaptive optics features created by the deformable mirror are masked and not part of the optimisation area, diffraction patterns from the spiders are only partially removed by the RDI reduction techniques. RDI creates in some region over-subtraction and the forward-modelling approach supposed to account for this effect is degenerate. 

\section{Incompatibility of the silicate, amorphous carbon and water ice porous particle model}
\label{app_si_ac_ice}

\subsection{Model description and results}
\label{subapp_si_ac_ice}

We parameterised the composition of such a dust model by the porosity without ice $P$, a fraction of vacuum removed by the ice $p_\text{H2O}$, and a silicate over organic refractory volume fraction $q_\text{Sior}$. The size of the smallest particles is written $s_\text{min}$ and the size distribution follows a differential power-law of exponent $\nu$.  The grid of these five free parameters is described in Table \ref{tab_grid}. 

\begin{table}
\caption{Grid of parameters for the 13\,000 models generated.}             % title of Table
\label{tab_grid}      % is used to refer this table in the text
\centering                          % used for centering table
\begin{tabular}{p{1.7cm} p{1cm} p{1cm} p{1cm} p{1.3cm}}        % centered columns (4 columns)
\hline\hline                 % inserts double horizontal lines
Parameter & Min. value & Max. value & $N_\text{sample}$ & Sampling  \\    % table heading 
\hline    
Scattering theory &  \multicolumn{2}{c}{Mie / DHS} & / & / \\   % inserts single horizontal line
 $s_{min}$ (\micron) & 0.1 & 100 & 13 & log. \\      % inserting body of the table
$p_{H_2O}$ (\%) & 1 & 90 & 5 & log. \\      % inserting body of the table
$P$ (\%) & 0 & 80 & 5 & linear \\      % inserting body of the table
$q_\text{Sior}$ & 0 & $\infty$ & 5 & $[0,0.5,1,2,\infty]$ \\
$\nu$ & -2.5 & -5.5 & 4 & linear \\
\hline                                   %inserts single line
\end{tabular}
\end{table}

For each model, we define a goodness of fit using a reduced chi squared to measure the distance between the model and our measurements, for the polarised fraction ($\chi^2_\text{pf}$), for the total intensity SPF ($\chi^2_\text{SPF}$), and overall ($\chi^2_\text{pf+SPF}=\chi^2_\text{pf} + \chi^2_\text{SPF}$). The parameters and goodness of fit of the best models are shown in Table \ref{tab_chi2_Mie_DHS} and their scattering properties are compared to our measurements in Fig. \ref{fig_marginal_pdf_Mie}. As also demonstrated in Sect. \ref{sec_optical_index_search}, we note that the polarised fraction is much more constraining than the SPF alone, due to the required scaling of the SPF: there are only five model with $\chi^2_\text{pf}<10$ while there are 1912 models with $\chi^2_\text{SPF}<10$. There is no model able to correctly match both the polarised fraction and the SPF: all models compatible with the polarisation fraction have a SPF steeper than observed. This tension was already revealed in \citet{Lebreton2012} and \citet{Schneider2006}, where the models compatible with the spectral energy distribution had an SPF steeper than observed. 

\begin{table}
%\begin{sidewaystable*}
\caption{Goodness of fit estimates and corresponding parameters for the best models with respect to the polarised fraction, total intensity SPF, or both.}            
\centering          
\label{tab_chi2_Mie_DHS}      
\begin{tabular}{ c | c c c }     % 7 columns 
\hline\hline       
  & best & best    & best  \\
  & polar. frac & SPF & combination \\
  \hline                  
Theory               & Mie   & Mie   & Mie \\
$\nu$                 & -3.5   & -3.5  & -4.5  \\
$q_\text{Sior}$  & 0       & 0.5    & $\infty$ \\
$p_{H_2O}$      & 0\tablefootmark{a}   & 0\tablefootmark{a}    & 0\tablefootmark{a} \\
$s_\text{min}$ ($\micron$)& 0.18   & 0.56  & 3.16 \\
P                       & 0\%   & 0\%   & 0\% \\
\hline                  
$\chi^2_\text{pf}$                     & \textbf{7.1} & 77.7                & 84.8          \\
$\chi^2_\text{SPF}$                      & 393             & \textbf{1.6}  & 395           \\
$\chi^2_\text{pf+SPF}$                     & 394             & 442               & \textbf{65.3} \\
\hline
\end{tabular}
\tablefoot{
\tablefoottext{a}{When the porosity without ice $P$ is $0$, there is no empty holes to be filled by water ice, hence no water ice in the composition and the fraction of vacuum removed by the ice $p_{H_2O}$ is unconstrained.}
}
\end{table}

\begin{figure}
    \centering
   \includegraphics[width=\hsize]{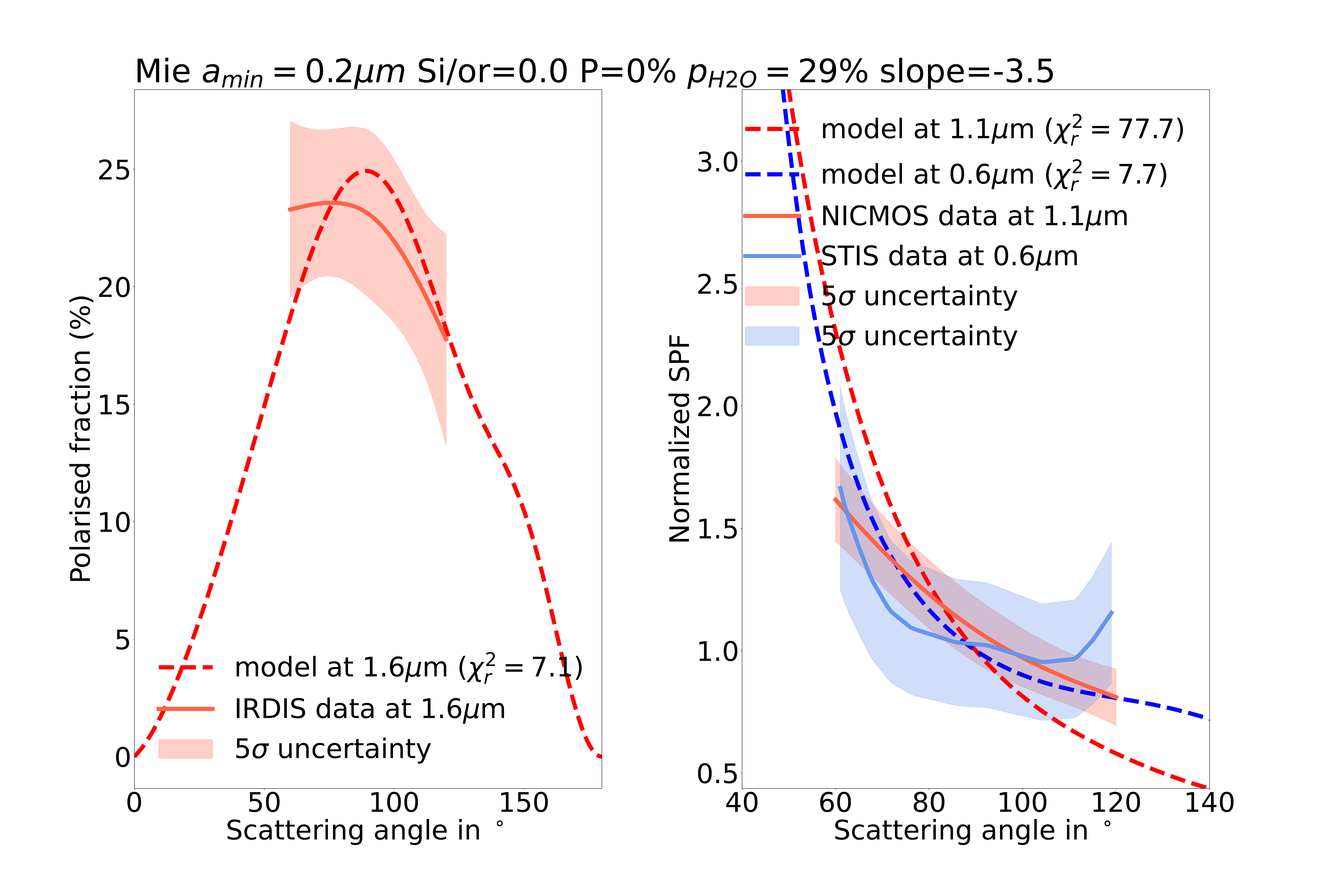}
   \includegraphics[width=\hsize]{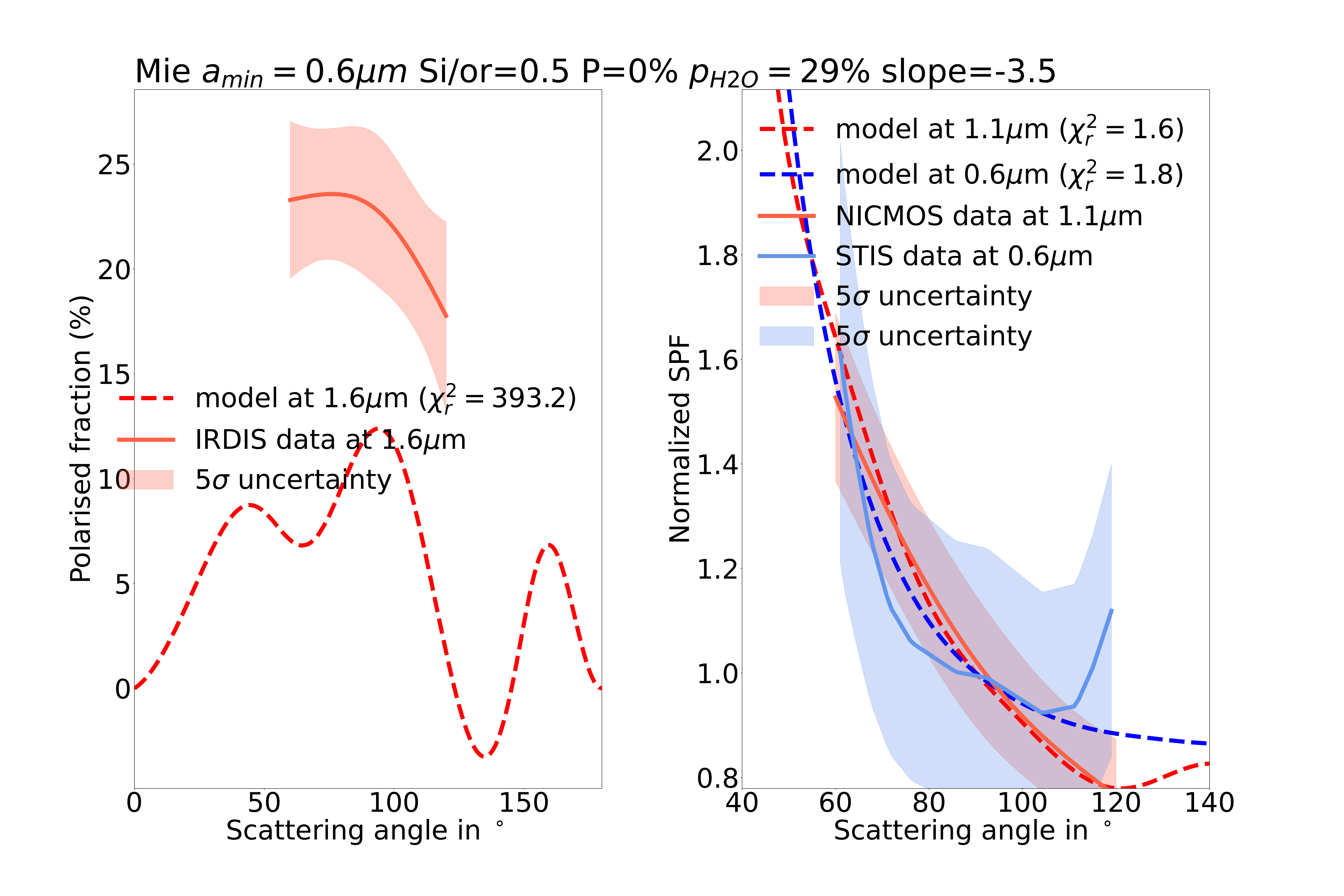}
   \includegraphics[width=\hsize]{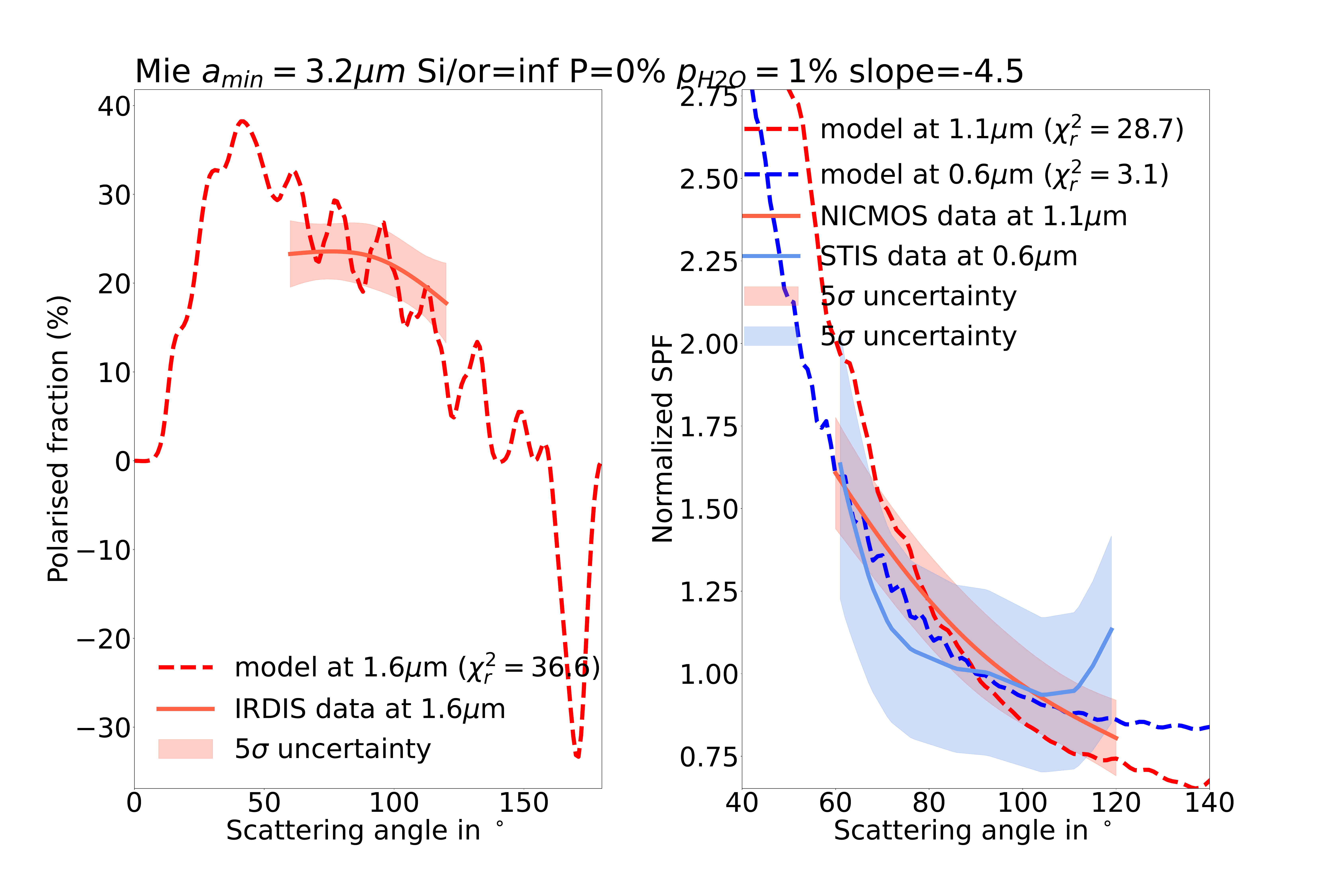}
   \caption{Comparison of the best models with the observations for the polarisation fraction (left) and total intensity SPF (right). The top, middle and bottom row shows respectively the best dust model for the polarisation fraction, for the NICMOS SPF and for the two observables simultaneously.} 
    \label{fig_model_observation_comp}
    \end{figure}

To provide an estimate of the range of acceptable models for each goodness of fit estimators, we carry out a Bayesian analysis and assign a probability $\Psi$ that the data are drawn from the model parameters. We do not have any a priori information on these parameters, we therefore assumed a uniform prior, corresponding to a uniform sampling of our parameters by our grid (see Table \ref{tab_grid}). The derivation of $\Psi$ follows the same methodology as described in Sect. \ref{sec_optical_index_search}.
Fig. \ref{fig_marginal_pdf_Mie} shows the inferred probability distribution for each of our five free parameters, after marginalisation against all other four parameters. It is shown here using the Mie theory because the best models are obtained with this theory. It highlights the absence of models compatible with both the polarised fraction and the SPF. The minimum particle size and the silicate to organic ratios favour indeed very different values whether we consider the polarised fraction alone (with pure organic sub-micron particles) or whether we consider also the SPF (favouring pure silicate micron-sized particles in this case). This discrepancy can be due to the inability of the Mie or DHS theory to capture the scattering properties of irregular or porous particles, or to the inadequacy of the dust composition made of only three basic constituents : amorphous carbon, silicates, ice. 

\begin{figure}
    \centering
   \includegraphics[width=\hsize]{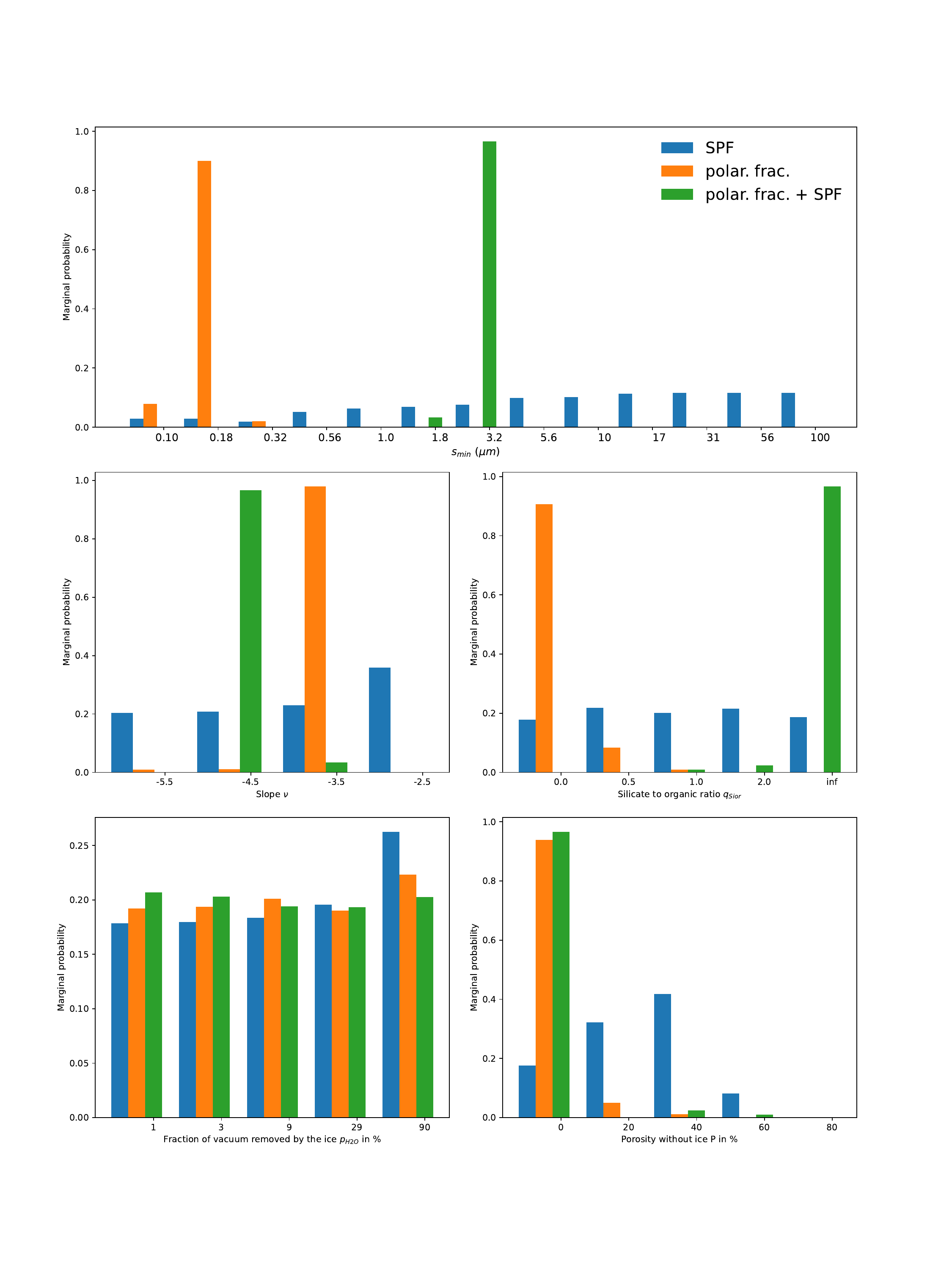}
   \caption{Marginal probability distributions of the five free parameters of our model, based on the polarised fraction (orange bars), on the NICMOS SPF (blue bars), or on the combination of both (green bars). These distributions were derived from models created using the Mie theory} 
    \label{fig_marginal_pdf_Mie}
    \end{figure}

\subsection{Validation of the assumption on the achromatic SPF between the J and the H band}
\label{subapp_assumption_validation}

When extracting the polarisation fraction from the IRDIS and NICMOS data, we made the assumption that the shape of the SPF in total intensity is the same between the IRDIS H band centred at $1.625\micron$ and the NICMOS wavelength centred at $1.104\micron$ (J band). The model developed in Appendix \ref{subapp_si_ac_ice} allows to test this assumption.

\begin{figure}
    \centering
   \includegraphics[width=\hsize]{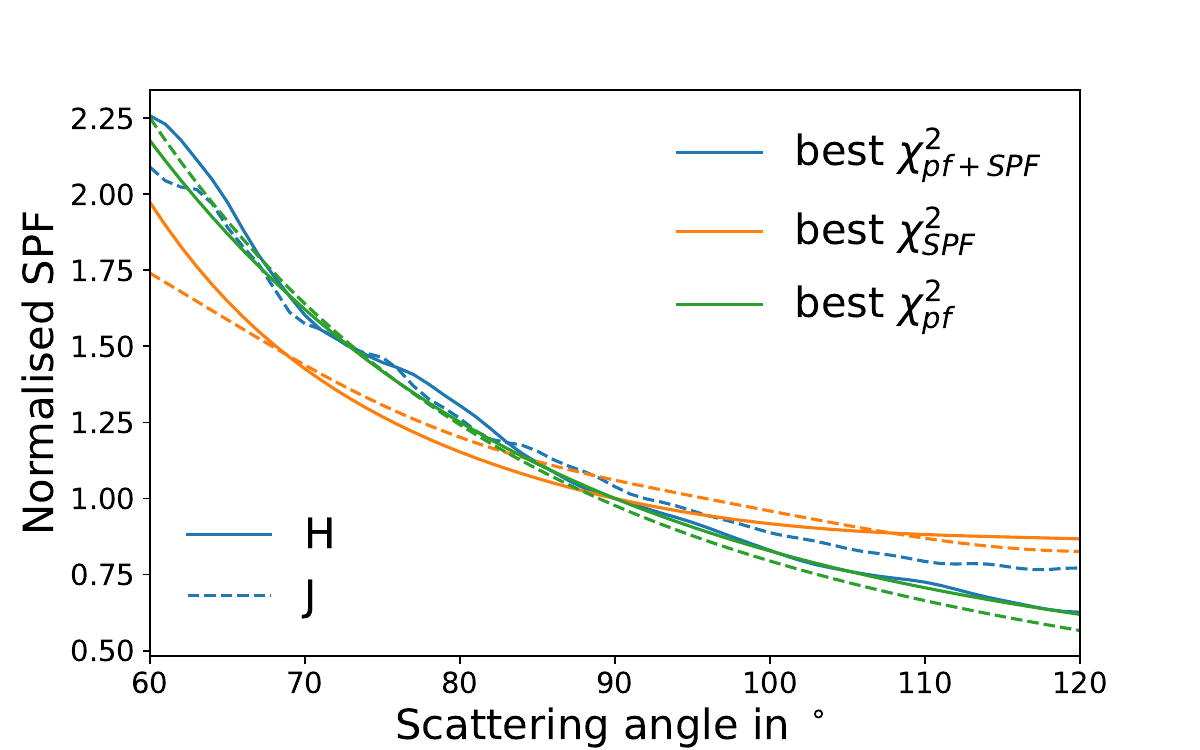}
   \caption{Comparison of the total intensity SPF in the H (plain lines) and J band (dashed lines). The comparison is shown for three dust models representing the best dust model for the polarisation fraction alone, for the NICMOS SPF alone or for the two observables simultaneously.}
    \label{fig_comparison_H_vs_J}
    \end{figure}

Among all the 6500 Mie models, the root mean square error (hereafter RMSE) between the SPF at $1.104\micron$ and $1.625um\micron$ is 0.85\% on average. It is less than 1\% in 81\% of the 6500 models, and less than 10\% in 99\% of them, with a maximum at 50\% for one specific model. For the best models matching either the polarised fraction alone, the SPF in total intensity alone, or both simultaneously (cf Fig. \ref{fig_model_observation_comp}, the RMSE amounts respectively to 0.12\%, 0.38\% and 0.57\%. The comparison between those SPF is shown in Fig. \ref{fig_comparison_H_vs_J}. We conclude that in most cases the differences between the SPF at J and H is smaller than 1\% and does not dominate our uncertainty.

\end{appendix}
 
\end{document}